\documentclass[11pt,a4paper]{article}
\pdfoutput=1
\usepackage{jheppub}

\usepackage{multirow,graphicx,amssymb,url,mathrsfs,amsmath}
\usepackage{wrapfig,boxedminipage,setspace,subfigure,epsfig}
\usepackage{amsxtra,amstext,latexsym,dsfont,amsfonts}
\usepackage[mathscr]{eucal}
\usepackage{color}
\usepackage[dvipsnames]{xcolor}
\usepackage{float}
\usepackage{slashed,comment}
\usepackage{kotex}
\usepackage{makeidx,pst-grad,pst-plot}
\usepackage{feynmf}

\usepackage{tikz}
\usetikzlibrary{calc,patterns,angles,quotes}
\usetikzlibrary{decorations.pathreplacing,decorations.markings,snakes}

\newcommand{\be}{\begin{equation}}
\newcommand{\ee}{\end{equation}}
\newcommand{\bea}{\begin{eqnarray}}
\newcommand{\eea}{\end{eqnarray}}

\title{Traversable wormholes via a double trace deformation involving $U(1)$ conserved current operators
}

\author[a]{Byoungjoon Ahn,}
\author[b]{Sang-Eon Bak,}
\author[a]{Viktor Jahnke,}
\author[a]{Keun-Young Kim}

\affiliation[a]{School of Physics and Chemistry, Gwangju Institute of Science and Technology, 123 Cheomdan-gwagiro, Gwangju 61005, Korea}
\affiliation[b]{Department of Physics, Arizona State University, AZ 85287, USA}

\emailAdd{bjahn123@gist.ac.kr}
\emailAdd{sbak2@asu.edu}
\emailAdd{viktorjahnke@gist.ac.kr}
\emailAdd{fortoe@gist.ac.kr}

\abstract{We study the effects of conservation laws on wormholes that are made traversable by a double trace deformation. After coupling the two asymptotic boundaries of a maximally extended $(d+1)$ dimensional black brane geometry with $U(1)$ conserved current operators, we find that the quantum matter stress-energy tensor of the corresponding bulk gauge fields in the hydrodynamic limit violates the averaged null energy condition (ANEC), rendering the wormhole traversable. Applying our results to axionic two-sided black hole solutions, we discuss how the wormhole opening depends on the charge diffusion constant, how this affects the amount of information that can be sent through the wormhole, and possible implications for many-body quantum teleportation protocols involving conserved current operators.
}

\makeatletter
\gdef\@fpheader{}
\makeatother
\begin{document}

\maketitle


\section{Introduction}

In holography \cite{Maldacena:1997re,Witten:1998qj,Gubser:1998bc}, maximally extended two-sided black holes are dual to a pair of conformal field theories entangled in a thermofield double state \cite{Maldacena_2003}. The wormhole connecting the two sides of the geometry can be made traversable by introducing a non-local coupling between the two asymptotic boundaries. This construction was first proposed by Gao, Jafferis and Wall \cite{Gao:2016bin}, and it allows us to send a message from one side of the geometry to the other. From the point of view of the boundary theory, this can be viewed as a teleportation protocol \cite{Gao:2016bin,Maldacena:2017axo}. 

In the simplest instance of quantum teleportation, two distant observers, often called Alice and Bob, share a pair of maximally entangled qubits. Alice has an additional qubit $|\psi\rangle$ that she wants to teleport to Bob. To do so, Alice performs a measurement on the two particles in her possession in a particular basis (the Bell basis) and reports the result to Bob through a classical communication channel. Finally, Bob performs a unitary operation in his qubit to obtain the desired quantum state \cite{PhysRevLett.70.1895}. The essential ingredients of this teleportation protocol are share entanglement, measurement, and classical communication, and it involves only three qubits.

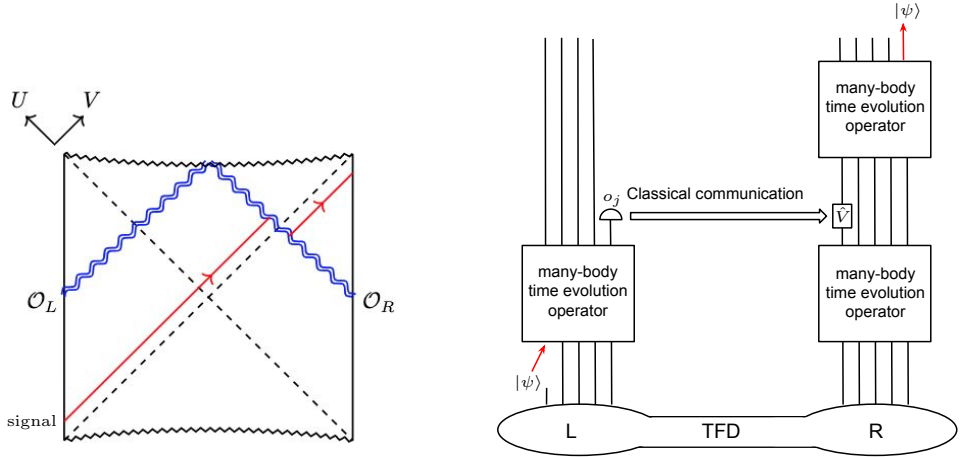
\begin{figure}[H]
    \centering
    \includegraphics[width=0.9\textwidth]{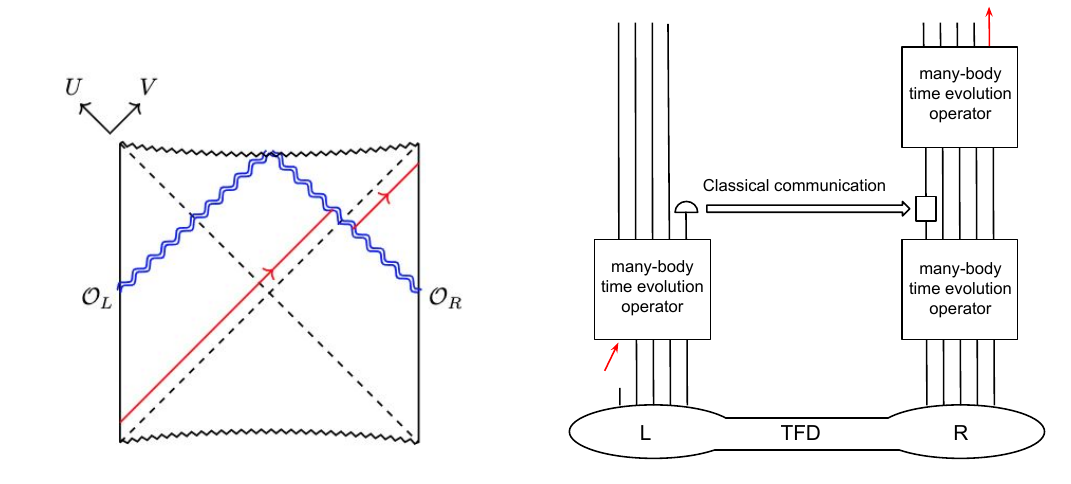}
    \put(-142,105){\tiny $o_j$}
    \put(-55,97){\tiny $\hat{V}$}
    \put(-175,35){\tiny $|\psi \rangle$}
    \put(-33,175){\tiny $|\psi \rangle$}
    \put(-365,20){\tiny signal}
    \caption{Left panel: Penrose diagram for the GJW traversable wormhole. The negative energy shock wave (shown in blue) causes a time advance $\Delta U$  for geodesics crossing it. The geodesics are actually continuous, but they appear to be discontinuous because the diagram is drawn using discontinuous coordinates. Using the equations of motion one can see that the time advance is proportional to the averaged null energy, i.e., $\Delta U \sim \int \langle T_{VV}\rangle dV $. The signal (shown in red) originated on the left boundary and propagating along the horizon ($U=0$) can cross to the other side of the geometry if $\Delta U <0$, which happens when ANEC is violated, i.e, $\int \langle T_{VV}\rangle dV <0$. Right panel: schematic representation of the traversable wormhole teleportation protocol proposed in~\cite{Schuster:2021uvg}. The quantum information (shown in red) is introduced in the left system at early times, gets scrambled with the other degrees of freedom, and reappears (unscrambles) in the right system at late times after a weak coupling between the two systems. The coupling is implemented as follows. We measure some operator $\mathcal{O}_L$ in the the left system, obtaining one of the possible values of $o_j$. We then apply the operator  $\hat{V} = e^{i \int h o_j \mathcal{O}_R } $ on the right system. For further details about this teleportation protocol, we refer to \cite{Schuster:2021uvg}.}
    \label{fig-TWprotocol}
\end{figure}

Gao, Jafferis and Wall (GJW) traversable wormhole is related to a new type of teleportation protocol, the so-called traversable wormhole teleportation protocol. It involves two copies of a strongly interacting many-body system entangled in a thermofield double state. Let us call them the left and the right systems, and consider the teleportation of a qubit from the left system to the right system. In this setup, the quantum information to be teleported is initially scrambled among the degrees of freedom of the left system. Then, after a weak coupling between the left and the right systems, the quantum information reappears (unscrambles) in the right system after a time of the order of the scrambling time. In this protocol, the thermofield double state plays the same role that the maximally entangled pair plays in conventional teleportation protocols, while measurement and classical communication are used to implement the coupling between the two systems. See Fig.~\ref{fig-TWprotocol}. The traversable wormhole protocol can be implemented in rather general chaotic many-body quantum systems, but the phenomenon of many-body quantum teleportation has distinct features in the case of systems that have an emergent gravitational description. This property makes the traversable wormhole protocol a powerful experimental tool to gain insights into the inner-working mechanisms of gauge-gravity duality \cite{brown2019quantum,Nezami:2021yaq,Schuster:2021uvg}.

Traversable wormholes violate the averaged Null Energy Condition (ANEC), which states that the integral of the stress energy tensor along complete achronal null geodesics is always non-negative
\be \label{eq-ANEC}
\int  T_{\mu \nu}  k^{\mu} k^{\nu} d \lambda \geq 0\,,
\ee
where $k^{\mu}$ is a tangent vector and $\lambda$ is an affine parameter. In classical theories, the construction of traversable wormholes is prevented by the Null Energy Condition (NEC) $T_{\mu \nu}  k^{\mu} k^{\nu} \geq 0$, which implies (\ref{eq-ANEC}) and is valid in physically reasonable theories. GJW construction of a traversable wormholes overcomes this difficulty by considering quantum mechanical effects.  They work in the context of the semi-classical approximation, in which the gravitational field is treated classically, but the matter fields are treated quantum mechanically. In this context, one writes Einstein's equations as follows
\be \label{eq-eom}
G_{\mu \nu} = 8 \pi G_N \langle T_{\mu \nu} \rangle\,,
\ee
where $G_{\mu \nu}$ is the Einstein tensor, and $\langle T_{\mu \nu} \rangle$ is the expectation value of the stress tensor in a given quantum state. Initially, before introducing the deformation, the wormhole connecting the two asymptotic boundaries is not traversable, which is consistent with the fact that the two boundary theories are not interacting. One then introduces a non-local deformation of the boundary theory
\be \label{eq-deformation-scalar}
 S_\text{bdry} \rightarrow S_\text{bdry} +  \int dt\, d^{d-1} x \, h(t,x)\, \mathcal{O}_L(-t,x) \mathcal{O}_R(t,x)\,,
\ee
which couples the two asymptotic boundaries. Here, $S_\text{bdry}$ denotes the action for the two copies of the boundary theory\footnote{In what follows, we will not need the explicit form of $S_\text{bdry}$, so we did not write it here.}, and $\mathcal{O}_{L,R}$ denotes a scalar operator that acts on the $d$-dimensional left/right boundary theory. For certain choices of $h(t,x)$, the deformation (\ref{eq-deformation-scalar}) gives rise to a stress energy tensor whose expectation value violates ANEC, 
\be \label{eq-ANECviolation}
\int  \langle T_{\mu \nu} \rangle  k^{\mu} k^{\nu} d \lambda < 0\,,
\ee
rendering the wormhole traversable. The physical picture is that the deformation (\ref{eq-deformation-scalar}) introduces negative energy in the bulk, whose backreaction is given in terms of a negative energy shock wave that causes a time advance for the geodesics crossing it, as opposed to the usual time delay caused by positive-energy shock waves. In this way, a signal originated on the left boundary can cross the wormhole and reach the right boundary after interacting with the negative energy, as shown in Fig.~\ref{fig-TWprotocol}. The ANEC can be violated in the GJW setup because the geodesics crossing the wormhole are not achronal -- early and late times points along the horizon can be connected by a timelike curve passing through the directly coupled boundaries.

GJW construction of a traversable wormhole involves a two-sided BTZ black hole, but their construction can be extended to several other gravitational setups, including 2-dimensional black holes \cite{Maldacena:2017axo, Bak:2018txn, Bak:2019mjd, Bak:2019nnu}, rotating BTZ black holes \cite{Caceres:2018ehr}, asymptotically flat black holes \cite{Fu:2019vco}, higher-dimensional hyperbolic black holes \cite{Ahn:2020csv}, and near-extremal traversable wormholes \cite{Fallows:2020ugr}. Other interesting developments include \cite{Almheiri:2018ijj,Couch:2019zni,Freivogel_2020, Geng:2020kxh,Nosaka:2020nuk,Levine:2020upy, Marolf:2019ojx, AlBalushi:2020kso, Emparan:2020ldj,Bao:2018msr,Hirano:2019ugo,Garcia-Garcia:2019poj,Numasawa:2020sty,Maldacena:2018mp,Fu_2019,Anand_2020,Anand_2022}. For a recent review of wormholes and their applications in gauge-gravity duality, see \cite{Kundu_2022}.

Despite the existence of a growing literature about wormholes that become traversable by a double trace deformation, most works only considered deformations involving scalar operators. It is then natural to question if such a construction is still possible for deformations involving other types of operators, such as vector and tensor operators. In particular, conserved current operators have different scrambling properties as compared to (non-conserved) scalar operators, and this is expected to affect the traversability properties of the wormhole.

In this work, we study the boundary deformations involving $U(1)$ conserved current operators. We are interested in the corresponding bulk gauge fields in the hydrodynamic limit. In this case, the gauge field displays a diffusive behavior which is expected to affect the traversability properties of the wormhole. In general, the diffusive behavior of bulk gauge fields leads to a power-law behavior of two-point functions \cite{Kovtun_2003,Caron-Huot:2009kyg} and out-of-time-order correlators\footnote{For a review on out-of-time-order correlators in holography, we refer to \cite{Viktor_2018}.} at late times ~\cite{Cheng:2021mop}. Since the wormhole opening can be computed as an integral involving a product of two-point functions, we expect it to display a power-law behavior at late times. Moreover, we expect the wormhole opening and consequently the bound on information transfer to depend on the transport properties of the black hole horizon. In particular, we would like to understand the effects of conservation laws on many-body traversable wormhole protocols and obtain the expected behavior for systems that admit a dual gravitational description. Is teleportation favored in this case? Or do the different scrambling properties of conserved currents result in a less efficient teleportation protocol? The benchmark holographic behavior is well understood in the case where the double trace deformation involves scalar field operators but has not yet been explored for deformations involving vector and tensor operators, and this might be relevant in the experimental realization of traversable wormhole teleportation protocols, especially because of this possible interplay between traversability and hydrodynamic behavior.

\subsection*{Organization of this work}
This work is organized as follows. In Sec.~\ref{sec-gravity}, we introduce our gravity setup. In Sec.~\ref{sec-scalarBC}, we review the boundary conditions for scalars fields associated with the GJW construction of a traversable wormhole. In Sec.~\ref{sec3}, we discuss boundary conditions for vector fields. In Sec.~\ref{section 4}, we consider a double trace deformation involving $U(1)$ conserved current operators and show that it leads to a violation of the averaged null energy condition. In Sec.~\ref{sec-disc}, we discuss our results. We relegate some technical details to the appendices \ref{sec-app0}, \ref{app-new} and \ref{app-A}.


\section{Gravity set-up} \label{sec-gravity}


We consider a general $(d+1)$-dimensional black brane background, with the line element of the form 
\be \label{eq-background}
ds^2= -G_{tt}(z) dt^2+G_{zz}(z) dz^2+G_{ij}(z) dx^j dx^j\,,
\ee
where $(t,x^i)$ are the boundary theory coordinates, with $i$ running from 1 to $d-1$, and $z$ is the AdS radial coordinate. We take the boundary to be located at $z=0$, where the geometry is assumed to asymptote AdS$_{d+1}$. We assume the horizon is located at $z=z_{h}$, where $G_{tt}$ has a first order zero and $G_{zz}$ has a first order pole. All the other metric components are assumed to be finite and non-zero at the horizon. For simplicity, we take $G_{ij} = \delta_{ij} G_{xx}$, which corresponds to assuming full rotational symmetry in the $x^i$ directions. Near the horizon, we write the metric functions $G_{tt}$ and $G_{zz}$ as
\be
G_{tt}= c_0 (z-z_h)\,,\,\,\,\,\,\, G_{zz}= \frac{c_1}{z-z_h}\,. 
\ee
With the above assumptions and by requiring regularity of the Euclidean continuation of the above line element at the horizon,
one obtains the inverse Hawking temperature as
\be
\beta = 4\pi \sqrt{\frac{c_1}{c_0}}\,.
\ee

\begin{figure}
\begin{center}
\begin{tabular}{cc}
\setlength{\unitlength}{1cm}
\hspace{0.1cm}
\includegraphics[width=5.7cm]{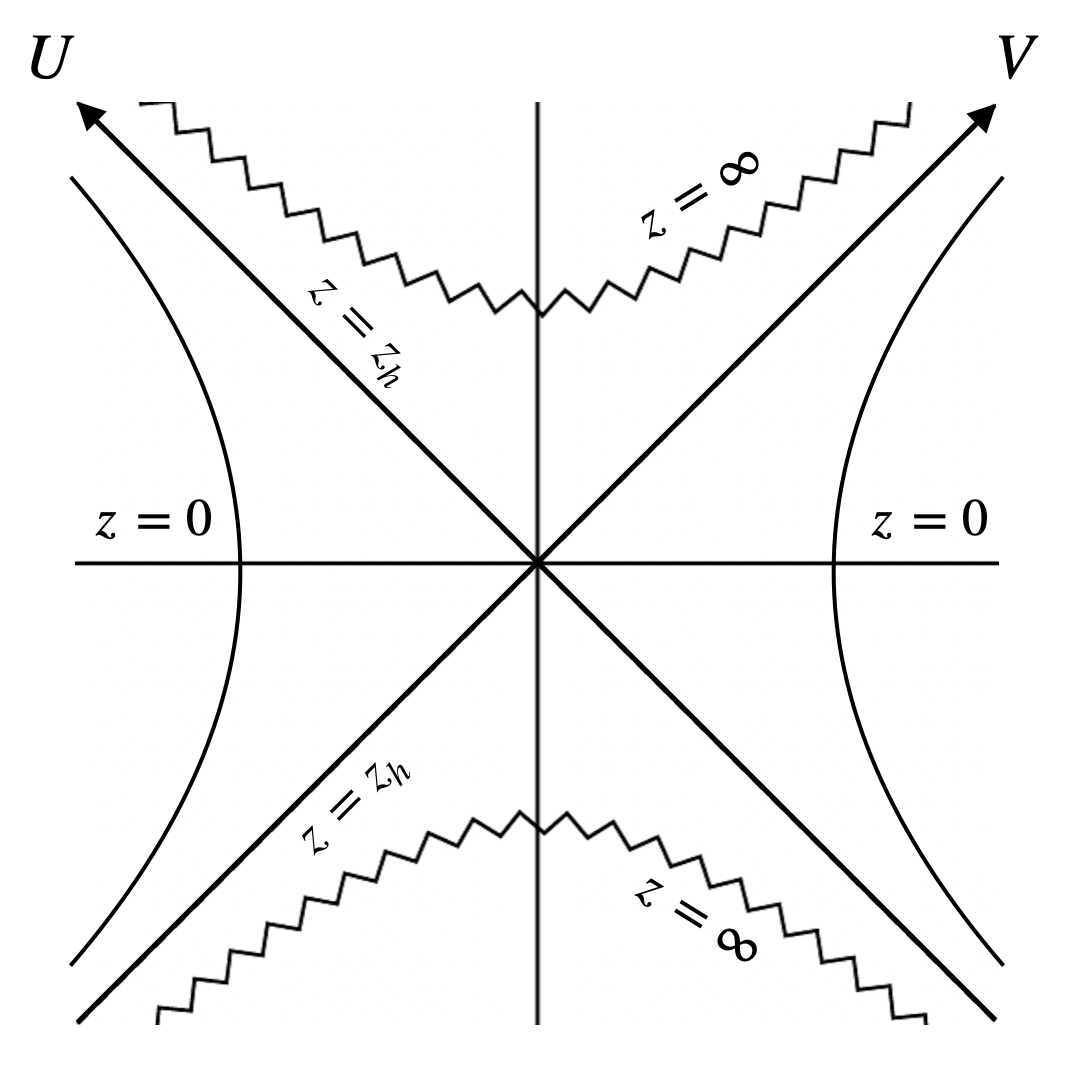}
\qquad &
\qquad
\includegraphics[width=6.8cm]{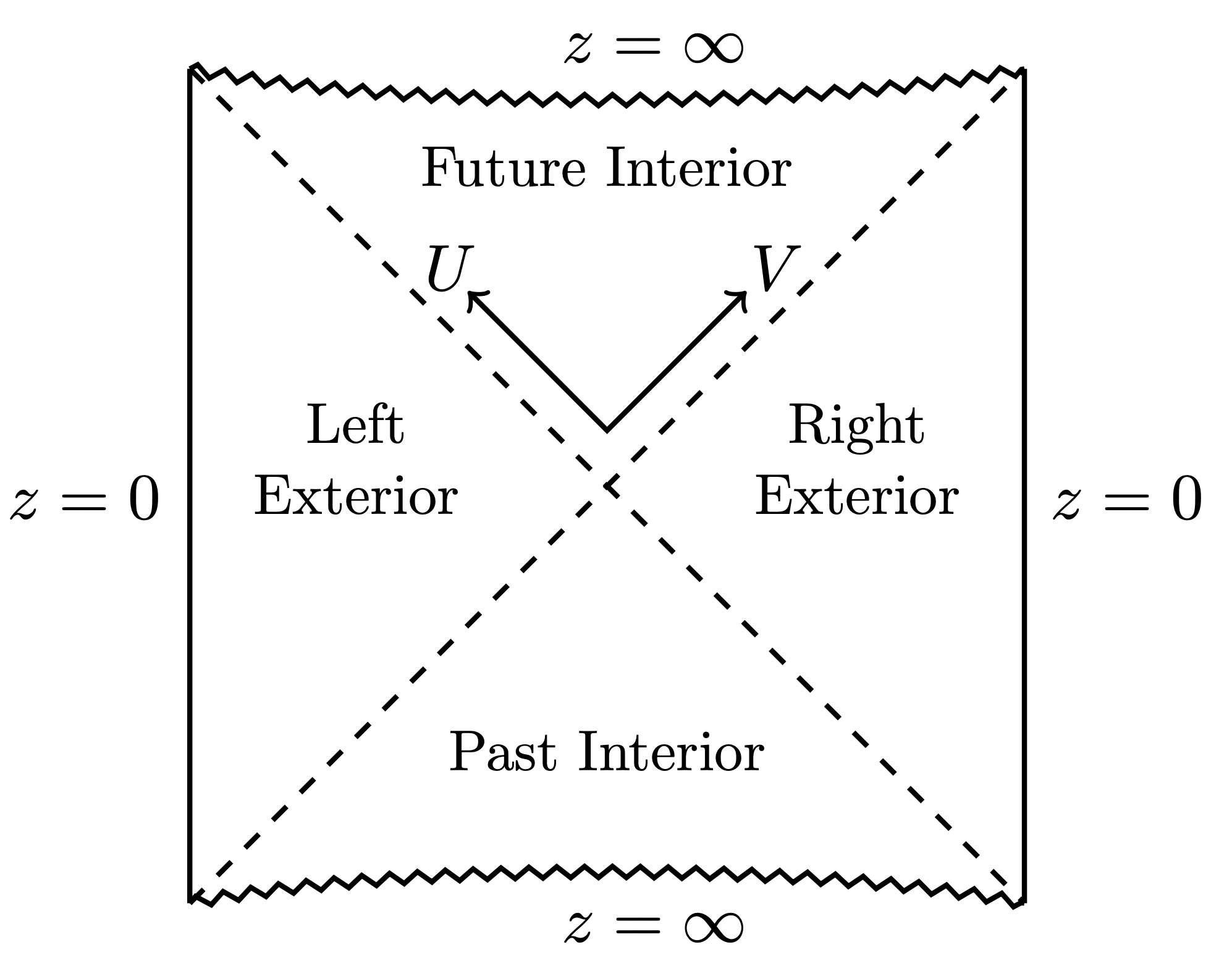}
\qquad
\end{tabular}
\end{center}
\caption{\small Kruskal diagram (Left) and Penrose diagram (Right)  for a two-sided asymptotically AdS black brane.}
    \label{fig-Penrose}
\end{figure}

To analyze the equations of motion near the horizon, it is convenient to define the tortoise coordinate 
\be\label{eq-tortoise}
dr =-\sqrt{\frac{G_{zz}}{G_{tt}}} dz
\ee
in terms of which the boundary is located at $r=0$ and the horizon at $r=-\infty$.
In terms of the tortoise coordinate, the line element reads
\be \label{eq-metric-tortoise}
ds^2=- G_{tt}(r)\left( dt^2-dr^2\right)+G_{xx}(r) \,\delta_{ij} \, dx^i dx^j\,.
\ee

Finally, to describe the globally extended spacetime, we introduce Kruskal-Szekeres coordinates $(U,V,x^i)$ as
\bea
U &=& -e^{-\frac{2 \pi}{\beta}(t-r)}, \,\,\, V = +e^{\frac{2\pi}{\beta}(t+r)}\,,\,\,\,\,\,\, \text{(right exterior region)} \nonumber\\
U &=& +e^{-\frac{2 \pi}{\beta}(t-r)}, \,\,\, V = -e^{\frac{2\pi}{\beta}(t+r)}\,,\,\,\,\,\,\, \text{(left exterior region)} \nonumber\\
U &=& +e^{-\frac{2 \pi}{\beta}(t-r)}, \,\,\, V = +e^{\frac{2\pi}{\beta}(t+r)}\,,\,\,\,\,\,\, \text{(future interior region)} \nonumber\\
U &=& -e^{-\frac{2 \pi}{\beta}(t-r)}, \,\,\, V = -e^{\frac{2\pi}{\beta}(t+r)}\,.\,\,\,\,\,\, \text{(past interior region)} \nonumber
\eea
In terms of these coordinates, the line element takes the form
\begin{align}
ds^2= G_{UV}(U V) dU dV+ G_{xx}(U V)\,\delta_{ij} dx^i dx^j\,, \qquad G_{UV}(U V) = -\frac{\beta^2}{4\pi^2} \frac{G_{tt}(UV)}{U V}\,.
\end{align}
where we used the fact that $r$ is a function of $UV$. In these coordinates the horizon is located at $U=0$ or at $V=0$. The left and right boundaries are located at $UV=-1$ and the past and future singularities at $UV=1$. The Kruskal and Penrose diagrams for this geometry are shown in Fig.~\ref{fig-Penrose}.


\section{Boundary conditions for scalar fields} \label{sec-scalarBC}
In this section, we review the boundary conditions for scalar fields that are associated with the GJW traversable wormhole. Let's consider a minimally coupled scalar field with mass $m$ propagating on the background (\ref{eq-background}). Near the boundary, the field behaves as
\be \label{eq-scalarNB}
\phi(z,y) = \alpha(y) z^{\Delta_-}+\beta(y) z^{\Delta_+}\,,
\ee
where $y=(t,x)$ denotes a boundary point, and
\be
\Delta_{\pm} = \frac{d}{2}\pm \nu\, \,\,\, \text{with}\,\,\, \nu = \sqrt{d^2/4+m^2}\,.
\ee
For $m^2 > -d^2/4+1$, the first term in (\ref{eq-scalarNB}) is non-normalizable and is associated to a deformation of the boundary theory of the form
\be
W_\alpha = \int d^dy\, j_{\alpha}(y)\, \mathcal{O}_{\alpha}(y)\, \,\,\,\, \text{with}\,\,\, \alpha(y)=j_{\alpha}(y)\,,
\ee
where the single trace operator $\mathcal{O}_{\alpha}(y)$ has scaling dimension $\Delta_+ = d/2+\nu$. 
The second term in (\ref{eq-scalarNB}) is normalizable, and is related to the expectation value of $\mathcal{O}_{\alpha}$
\be
 \beta(y) = 2 \nu \, \langle \mathcal{O}_{\alpha}(y)\rangle\,.
\ee
For $-d^2/4 < m^2 < -d^2/4+1$, both terms in (\ref{eq-scalarNB}) are normalizable. In this regime, we are free to impose boundary conditions on either $\alpha$ or $\beta$. Each choice corresponds to a different boundary theory. In the boundary theory in which $\alpha$($\beta$) is fixed, the bulk field $\phi$ is dual to an operator of dimension $\Delta_+$($\Delta_-$). We refer to these boundary theories as CFT${}_{\Delta_+}$ and CFT${}_{\Delta_-}$. In particular, the so-called {\it alternative boundary condition}
\be
\beta(y)= j_{\beta}(y)\,,
\ee
is associated to deformations of the form
\be
W_{\beta} = \int d^dy \, j_{\beta}(y)\mathcal{O}_{\beta}(y)\,,
\ee
in which the boundary operator $\mathcal{O}_{\beta}(y)$ has scaling dimension $\Delta_-=d/2-\nu$. Note that, since $-d^2/4 < m^2 < -d^2/4+1$, one has $\frac{d-2}{2} < \Delta_- < \frac{d}{2}$. In this case, the leading order term is now associated to the expectation value of  $\mathcal{O}_{\beta}(y)$:
\be
 \alpha(y) = - 2 \nu \, \langle \mathcal{O}_{\beta}(y)\rangle\,.
\ee

The above equations allow us to symbolically write the deformation as $W_{\beta} \sim \int d^dy\, \beta\, \alpha$. A linear boundary condition relating the faster and the slower falloff parts
\be \label{eq-linearBC}
\beta  = h\, \alpha\,, 
\ee
corresponds to a double trace deformation of the form~\cite{Witten:2001ua}
\be \label{eq-defbeta}
W_{\beta} \sim \int d^dy \,h\, \alpha^2 \sim \int d^dy \, h\,\mathcal{O}_{\beta}^2\,,
\ee
which is a relevant deformation, because $\Delta_- <d/2$. Starting from the CFT${}_{\Delta_-}$, the deformation (\ref{eq-defbeta}) produces a renormalization group flow which is expected to end at the CFT${}_{\Delta_+}$ in the infrared. We now explain how the linear boundary condition (\ref{eq-linearBC}) can be used to construct a traversable wormhole. 

GJW construction of a traversable wormhole relies on the use of a relevant deformation, because in this case the near boundary geometry is not modified by backreaction in an uncontrolled way \cite{Gao:2016bin}. In this way, their construction can be shown to be embeddable in a UV complete theory of gravity. To have a relevant deformation, they consider a massive scalar field with mass in the range $-d^2/4 < m^2 < -d^2/4+1$ with the alternative boundary condition, in which the boundary operator has dimension $\Delta_- = d/2-\nu$.
For the fields propagating in the left and right exterior regions, we expect the following near boundary behaviors
\begin{align}
\phi(z \rightarrow 0_L,y) &= \alpha_L(y) z^{\Delta_-}+\beta_L(y) z^{\Delta_+}\,, \\
\phi(z \rightarrow 0_R,y) &= \alpha_R(y) z^{\Delta_-}+\beta_R(y) z^{\Delta_+}\,.
\end{align}
In order to have a deformation coupling the two asymptotic boundaries, GJW impose the following boundary conditions
\be
\beta_R = h \, \alpha_L\,,\,\,\,\,\beta_L = h \, \alpha_R
\ee
which correspond to a non-local double trace deformation of the form
\be \label{eq-defLR}
W_{\beta} \sim \int d^dy \, h\,\alpha_L \, \alpha_R \sim \int d^dy \,h(y)\,\mathcal{O}^{(L)}_{\beta}(y) \, \mathcal{O}^{(R)}_{\beta}(y)\,.
\ee
The deformation (\ref{eq-defLR}) modifies the 1-loop expectation value of the scalar field stress-energy tensor. For some choices of $h(y)$, this leads to a violation of the ANEC that renders the wormhole traversable.


\section{Boundary conditions for vector fields}\label{sec3} 
In this section, we briefly review possible boundary conditions for vector fields in AdS/CFT. A general discussion about this topic was first presented in \cite{Marolf:2006nd}. 
We consider a massless vector field $A_{\mu}$ propagating on the background (\ref{eq-background}) with action
\be \label{eq-actionvector}
S=-\frac{1}{4}\int_{\mathcal{M}} d^{d+1}{\bf r} \sqrt{-g}\, F_{\mu \nu} F^{\mu \nu}+ \int_{\partial \mathcal{M}} d^{d}y \sqrt{-\gamma} \, n_{\mu} F^{\mu \nu} A_{\nu}\,,
\ee
where ${\bf r}=(z,t,x) \in \mathcal{M}$ denotes a bulk point and $y=(t,x) \in \partial \mathcal{M}$ denotes a boundary point. Here, $\gamma$ denotes the determinant of the induced metric on the boundary, and $n_{\mu}$ is the outward pointing unit vector normal to $\partial \mathcal{M}$. 
The equation of motion for $A_{\mu}$ reads
\be \label{eq-EOM}
\nabla_{\nu} F^{\mu \nu} =0\,.
\ee
\subsection{Dirichlet boundary conditions for vector fields}
Near the boundary ($z \rightarrow 0$), the gauge field behaves as
\be
A_\mu(z \rightarrow 0) = a_\mu + b_\mu \, z^{d-2} + \dots
\ee
In general, when $d \geq 4$, the leading term $a_\mu$ is non-normalizable and must be fixed. This corresponds to imposing a Dirichlet boundary condition on $A_\mu$. From the point of view of the boundary theory, such boundary condition correspond to a deformation by a global $U(1)$ conserved current operator
\be
W_{A} = \int_{\partial \mathcal{M}} d^dy \,  a_\mu\,J^\mu\,,
\ee
with $a_\mu$ acting as a source for $J^\mu$. Note that, since $A_{\mu} = a_\mu$ at the boundary, we can write $W_{A} = \int_{\partial \mathcal{M}} d^dy \,  A_\mu\,J^\mu$. The expectation value of the conserved current is given by 
\be
\langle J_{\mu} \rangle \sim b_\mu\,.
\ee
This allow us to write $W_{A} \sim \int_{\partial \mathcal{M}} d^dy \,  a^\mu\,b_\mu\,$. Similarly to the scalar field case, a double trace deformation can be introduced by imposing a linear relation between $a^\mu$ and $b_\mu$, i.e., $a^\mu = Q^{\mu \nu} b_\nu$, which leads to 
\be \label{eq-Wa}
W_A \sim \int_{\partial \mathcal{M}} d^dy \, Q^{\mu \nu}\, b_\mu\,b_\nu \sim \int_{\partial \mathcal{M}} d^dy \, Q^{\mu \nu}\, J_\mu\,J_\nu \, ,
\ee
which is an irrelevant deformation, with dimension $2d-2$, because the scaling dimension of $J^\mu$ is $d-1$. Here $Q^{\mu \nu}$ is a set of parameters that we will specify later.

Just like in the scalar field case, one can construct a traversable wormhole by coupling the two asymptotic boundaries of an asymptotically AdS two-sided black hole with conserved current operators. This can be done as follows. First, we write the near boundary behavior of the gauge field as follows
\bea
A_\mu(z \rightarrow 0_R) &= a^{(R)}_\mu + b^{(R)}_\mu \, z^{d-2} + \dots\\
A_\mu(z \rightarrow 0_L) &= a^{(L)}_\mu + b^{(L)}_\mu \, z^{d-2} + \dots
\eea
where the superscripts $R$ and $L$ denote the right anf left exterior regions, respectively. Then, an irrelevant double trace deformation can be introduced by imposing the following boundary condition
\be
a^{\mu (R)} = Q^{\mu \nu} b^{(L)}_{\nu}\,, \,\,\,\,\,\, a^{\mu (L)} = Q^{\mu \nu} b^{(R)}_{\nu}\,,
\ee
which leads to a deformation of the form
\be \label{eq-Wa2}
W_A \sim \int_{\partial \mathcal{M}} d^dy \, Q^{\mu \nu}\, J^{(L)}_\mu\,J^{(R)}_\nu\,.
\ee
We show in the next section that for some choices of $Q^{\mu \nu}$ the deformation (\ref{eq-Wa2}) leads to a violation of the ANEC, rendering the wormhole traversable. 

In this work, we consider the deformation (\ref{eq-Wa2}) because we are interested in the hydrodynamic limit of the gauge fields, associated with long-distance physics. We can think about our setup as the IR fixed point of a renormalization group (RG) flow. As explained in the next section, a relevant deformation can be introduced if one considers gauge fields satisfying Neumann boundary conditions. In this case, the initial UV theory flows under the RG flow to an IR fixed point in which the gauge fields respect Dirichlet boundary conditions, which corresponds to our case. By that reasoning, we expect that our construction has a well-defined UV fixed point. 

Our analysis should be contrasted with the one conducted by performed by Gao, Jafferis, and Wall \cite{Gao:2016bin}. In their case, from the dual CFT viewpoint, the double trace deformation is relevant in the context of the renormalization group, leading to a well-defined UV fixed point. The deformation we employ in our analysis (\ref{eq-Wa2}), however, is irrelevant, and thus, there is no guarantee that it leads to a well-defined fixed point. Nevertheless, as discussed in \cite{Marolf:2006nd}, there is evidence supporting that the deformation (\ref{eq-Wa2}) results in a well-defined UV fixed point in four, five, and six bulk spacetime dimensions.

\subsection{Neumann boundary conditions for vector fields}
In the previous section, we adopted the most common boundary condition for vector fields in AdS/CFT, which corresponds to imposing Dirichlet boundary conditions on $A_\nu$. As pointed out in \cite{Marolf:2006nd}, for $d=3,4$ and $5$, it is possible to consider an alternative boundary condition for vector fields, in which one imposes Neumann boundary condition on the gauge field by fixing
\be
F^{\nu} = \sqrt{-\gamma} \, n_{\mu} F^{\mu \nu} \,,
\ee
at the boundary. This boundary condition fixes the value $b_{\mu}$, but leaves $a_{\mu}$ unconstrained. For $d=3$, this corresponds to a deformation of the boundary theory by an operator $\mathcal{O}_{F^{\nu}}$, with dimension one. One can show that this operator is a $U(1)$ gauge field \cite{Witten:2003ya,Leigh:2003ez, Marolf:2006nd}.

Imposing the condition $b^\mu = Q^{\mu \nu} a_\nu$ corresponds to introducing a relevant double trace deformation of the form \cite{Marolf:2006nd}
\be \label{eq-defF}
W_F = \int_{\partial \mathcal{M}} d^d y \, Q^{\mu \nu} \, \mathcal{O}_{F^\mu}\, \mathcal{O}_{F^\nu}\,,
\ee
which has dimension 2. Under the deformation (\ref{eq-defF}), the boundary theory associated to Neumann boundary conditions flows in the infrared to the boundary theory associated to Dirichlet boundary conditions \cite{Marolf:2006nd}.

Once again, an wormhole connecting the two sides of an asymptotically AdS two-sided black hole can be made traversable by introducing the following relevant deformation
\be 
W_F = \int_{\partial \mathcal{M}} d^d y \, Q^{\mu \nu} \,  \mathcal{O}^{(L)}_{F^{\mu}}\, \mathcal{O}^{(R)}_{F^\nu}\,,
\ee
which is associated to boundary conditions of the form 
\be
b^\mu_{(R)} = Q^{\mu \nu} a^{(L)}_\nu\,,\,\,\,\, b^\mu_{(L)} = Q^{\mu \nu} a^{(R)}_\nu.
\ee


\section{Opening the wormhole with conserved current operators}\label{section 4}

In this section, we propose a generalization of the GJW traversable wormhole that involves a double trace deformation constructed out of conserved currents
\be \label{eq-deformation}
\delta H(t) = \int d^{d-1} x \, Q^{\mu \nu}(t,x) J_{\mu}^{(L)}(-t,x)J_{\nu}^{(R)}(t, x)\,.
\ee
For some choices of $Q^{\mu \nu}$, the deformation (\ref{eq-deformation}) introduces negative null energy in the bulk, which makes the wormhole traversable. 
For simplicity, we consider the coupling between the two conformal field theories as involving only $J^{(R)}_t$ and $J^{(L)}_t$, i.e., we set $Q^{\mu \nu}$ as
\be\label{perturbationJ0}
Q^{\mu \nu}(t_1,x_1) = \delta^{\mu}_t\, \delta^{\nu}_t\, h(t_1,x_1)\,.
\ee
Here, $h(t_1,x_1)$ controls the coupling between the two boundary theories. We choose a perturbation that is instantaneous in time $t_0$ and uniform in the transverse space
\be \label{eq-h}
h(t_1,x_1)= h \,\delta \left( \frac{2\pi}{\beta}(t_1-t_0) \right) = h\, V_0 \delta (V_1-V_0)\,,
\ee
where $h$ is a constant. By dimensional analysis, $h$ has dimensions of $[E]^{2-d}$. Later, we will factor out this dimensional dependence and write this constant as $h = \tilde{h}\, T^{2-d}$.

The traversability of the wormhole can be measured near the event horizon $(U=0)$ by the averaged null energy
\be
\mathcal{A}= \int dV T_{VV}\,,
\ee
where $T_{VV}$ is the $VV$ component of the Maxwell stress energy tensor. To study the effect of the deformation on $T_{VV}$, we compute the 1-loop Maxwell stress energy tensor using a point splitting technique.
The Maxwell stress energy tensor is given by
\be
T_{\mu \nu}=- F_{\alpha \mu} F^{\alpha}_{\nu}+\frac{1}{4}g_{\mu \nu} F_{\alpha \beta}F^{\alpha \beta}\,.
\ee
To simplify our calculation, we consider that only $A_V$ and $A_{U}$ are non-zero, and we consider a metric of the form $ds^2=G_{UV}dU dV+ G_{ij} dx^i dx^j$. With these assumptions, the $VV$ component of the stress energy tensor reads
\be
T_{VV} = - F_{\alpha V} F^{\alpha}_{V}= - F_{i V} F^{i}_{V} = - G^{ij} F_{i V} F_{j V}\,.
\ee
Now, using $F_{i V}=\partial_i A_V-\partial_V A_i =\partial_i A_V$ and the point splitting method, we write
\begin{align} \label{eq-Tvv}
\langle T_{VV} \rangle =& - \lim_{{\bf r}'\rightarrow {\bf r}} G^{ij} \langle  \partial_i A_V({\bf r})  \partial_{j'} A_{V}({\bf r}') \rangle \\
=& - \lim_{{\bf r}'\rightarrow {\bf r}} G^{ij} \partial_i \partial_{j'}  \langle   A_V ({\bf r})  A_{V} ({\bf r}') \rangle, \nonumber
\end{align}
where $\langle   A_V ({\bf r})  A_{V} ({\bf r}') \rangle$ 
is the bulk two-point function between two points ${\bf r}=(z,t,x)$, and ${\bf r'}=(z',t',x')$\footnote{Note that $\langle T_{VV} \rangle$ can be written in terms of the gauge invariant quantity $F_{iV}= \partial_i A_V - \partial_V A_i =\partial_i A_V$, i.e, $\langle T_{VV} \rangle =\lim_{{\bf r}'\rightarrow {\bf r}} g^{ij} \langle F_{iV}({\bf r}) F_{jV}({\bf r}') \rangle $. This makes it clear that the expression is gauge invariant, as expected on physical grounds.}.

\subsection{Modified bulk two-point function}
To compute stress energy tensor involving the deformation $\delta H$ in (\ref{eq-deformation}), we first compute the modified bulk two-point function. Due to the deformation, the gauge field operators in (\ref{eq-Tvv}) evolves in time with the operator $U(t,t_0) = \mathcal{T} e^{-i \int_{t_0}^{t} dt_1 \delta H(t_1)}$ in the interaction picture. Then, the modified bulk two-point function in (\ref{eq-Tvv}) is given by
\begin{align}
\mathcal{G}_{VV}({\bf r};{\bf r}')
&\equiv  \langle A_{V}^{(H)}({\bf r}) A_{V}^{(H)}({\bf r}') \rangle \nonumber \\
&= \langle U(t,t_0)^{-1}A_V^{(I)}({\bf r})U(t,t_0) U(t',t_0)^{-1}A_{V}^{(I)}({\bf r}')U(t',t_0) \rangle \,,
\end{align}
where we used the superscripts $H$ and $I$ to distinguish between the Heisenberg and interaction picture, respectively. 
For notational simplicity, we will omit the superscript $I$ in what follows.

Working at first order in perturbation theory, we expand 
\begin{align}
   U(t,t_0) & = \mathcal{T} e^{-i \int_{t_0}^{t} dt_1 \delta H(t_1) } = 1 -i \int_{t_0}^{t} dt_1 \delta H(t_1)\,, \\
& \mathcal{G}_{VV}({\bf r};{\bf r}')= \mathcal{G}_{VV}^{(0)}({\bf r};{\bf r}')+ \mathcal{G}_{VV}^{(1)}({\bf r};{\bf r}')\,, 
\end{align}
where $\mathcal{G}_{VV}^{(0)}({\bf r};{\bf r}')$ is the unperturbed two-point function of $A_V$.
The first correction to the two-point function of $A_V$ is obtained as follows,
\begin{align}
   \mathcal{G}_{VV}^{(1)}({\bf r};{\bf r}') &=  i \int_{t_0}^{t'} dt_1 \langle A_V({\bf r}) [\delta H(t_1),A_{V}({\bf r}')] \rangle + i \int_{t_0}^{t} dt_1  \langle  [\delta H(t_1),A_V({\bf r})] A_{V}({\bf r}') \rangle\, \\
   &= i \int_{t_0}^{t'} dt_1\, d^{d-1} x_1 Q^{\mu \nu}  \langle A_V({\bf r}) [ J_{\mu}^{(L)}(-t_1, x_1) J^{(R)}_{\nu}(t_1, x_1),A_{V}({\bf r}')] \rangle +({\bf r} \leftrightarrow {\bf r'}) \nonumber \\
     &= i \int_{t_0}^{t'} dt_1\,d^{d-1} x_1 Q^{\mu \nu} \langle A_{V}({\bf r}') J^{(L)}_{\mu}(-t_1,x_1) \rangle \langle [J^{(R)}_{\nu}(t_1,x_1),A_V({\bf r})] \rangle+({\bf r} \leftrightarrow {\bf r'})\nonumber\,\,. \nonumber
\end{align}
In the second line, we put \eqref{eq-deformation} into the expression. In the third line, we assume that the fields $A_V({\bf r})$ and $A_{V}({\bf r}')$ act on the right exterior region. Then, we can use large $N$ factorization and the fact that the operator on the right exterior region commutes with the operators on the left boundary.

Let us introduce the Wightman and retarded bulk-to-boundary propagators for the right exterior region as
\bea
K^{\text{w}}_{\mu \nu}(z,t,x;t_1,x_1)&\equiv& -i \langle A_\mu(z,t,x) J^{(R)}_\nu(t_1,x_1) \rangle\,, \label{eq-Kw}\\
K^\text{ret}_{\mu \nu}(z,t,x;t_1,x_1)&\equiv& -i \langle [A_\mu(z,t,x),J^{(R)}_{\nu}(t_1,x_1)] \rangle \label{eq-Kr} \,.
\eea
for $t>t_1$. Then, the expression of the first order correction to the bulk two-point function of $A_V$ can be rewritten as follows,
\begin{align}
 \mathcal{G}_{VV}^{(1)}({\bf r};{\bf r}') &= i \,\int_{t_0}^{t'} dt_1\, d^{d-1}x_1\,Q^{\mu \nu} \, K^\text{w}_{V \mu}(z',t', x' ; -t_1-\frac{i \beta}{2}, x_1) K^\text{ret}_{V \nu}(z,t,x ; t_1, x_1)+(t' \leftrightarrow t) \, \nonumber\\
 &= i \,\int_{t_0}^{t'} dt_1\, d^{d-1}x_1\,h(t_1, x_1) \, K^\text{w}_{V t}(z',t',x' ; -t_1- \frac{i \beta}{2}, x_1) K^\text{ret}_{V t}(z,t, x; t_1, x_1)+(t' \leftrightarrow t) \,. \nonumber
\end{align}
In the first line, we used the KMS condition \cite{Haag:1967sg} which gives the relation between $J^{(L)}_{\mu}$ and $J^{(R)}_{\mu}$, and in the second line, we included the perturbation \eqref{perturbationJ0}.

In this work, we consider the bulk-to-boundary propagators (\ref{eq-Kw}) and (\ref{eq-Kr}) under the hydrodynamic approximation because in that case the gauge field reveals a diffusive behavior, controlled by a charge diffusion constant. In this approximation, we focus on large times and large distances. This translates to solving our equations of motion for low-energy modes with large wavelengths. The bulk-to-boundary propagators in the hydrodynamic limit can be written as (see Appendix~\ref{app-A} or \cite{Cheng:2021mop})
\bea
 K^\text{ret}_{Vt}(V,x;t_1,x_1)&=& 2 i \, T \,D_c \,\theta(V-e^{2\pi T t_1}) \,\,\partial_V \int \frac{d^{d-1} k}{(2\pi)^{d-1}} \frac{ e^{i k \cdot (x-x_1)}}{\left( V e^{-2\pi T t_1}-1 \right)^{\frac{D_c k^2}{2 \pi T}}} \,, \\
 K^\text{w}_{Vt}(V, x;t_1, x_1)  &=& \,T\, D_c\,  \partial_V \int \frac{d^{d-1} k}{(2\pi)^{d-1}} \frac{1}{2 \sin \left( \frac{D_c k^2}{2 T} \right)} \frac{e^{i  k\cdot  (x-x_1)} }{\left( 1-V\, e^{-2\pi T t_1} \right)^{\frac{D_c k^2}{  2\pi T}}} \,,
\eea
where $D_c$ is the charge diffusion constant:
\be \label{eq-D}
D_c= \frac{\sqrt{-G}}{G_{xx}\sqrt{G_{tt}G_{zz}}}\Big|_{z_h} \int_{0}^{z_h} dz \frac{G_{tt}G_{zz}}{\sqrt{-G}}\,.
\ee

Using the above expressions, $\mathcal{G}_{VV}^{(1)}$ is obtained as 
\begin{align}
\mathcal{G}_{VV}^{(1)}= &-2 T^2 D_c^2 \int_{V_0}^{V'}\frac{dV_1}{2 \pi T V_1} h(V_1) \int \frac{d^{d-1}k}{(2\pi)^{d-1}} \frac{ D_k^2}{2 \sin (\pi D_k) }\frac{e^{i k \cdot(x-x')}}{(V/V_1-1)^{D_k+1}(1+ V_1 V')^{D_k+1}} \nonumber\\
&+(V \leftrightarrow V') \,,
\end{align}
where $D_k=\frac{D_c}{2\pi T} k^2$, and $V_1 = e^{2\pi T t_1}$. 

\subsection{1-loop stress tensor}
\begin{figure}[H]
    \centering
    \includegraphics[width=0.55\textwidth]{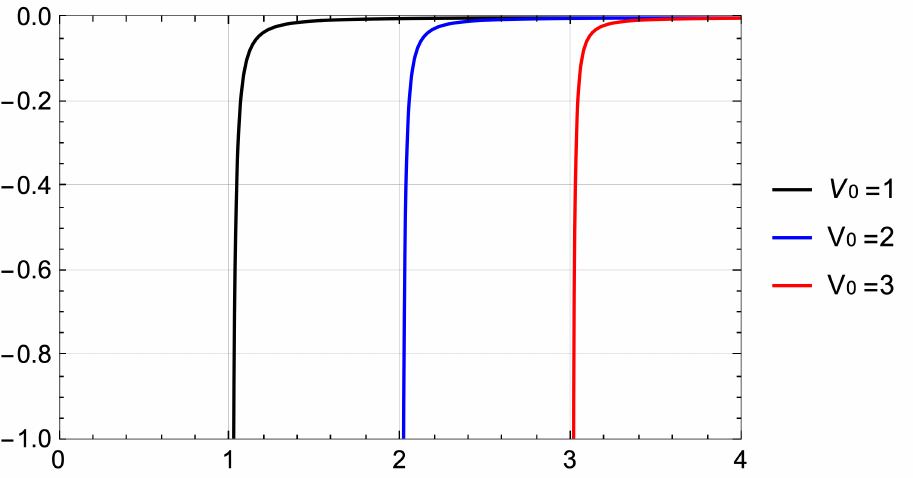}
    \put(-130,-7){$V$}
    \put(-255,40){\rotatebox{90}{$ \mathcal{N}^{-1}\langle T_{VV} \rangle$}}
    \caption{ Stress-energy tensor (\ref{eq-Tvv-num}) versus $V$ for several values of the insertion time $V_0$. Here, we set $\tilde{h}=-1$, $d=4$, and $u_\text{max}=1/2$.}
    \label{fig-Tvv3}
\end{figure}

The $VV$ component of the 1-loop expectation value can then be computed as 
\be\label{TVVintegrand1}
 \langle T_{VV} \rangle = - \lim_{{\bf r}'\rightarrow {\bf r}} G^{ij}\partial_i \partial_{j'} \mathcal{G}_{VV}^{(1)}({\bf r};{\bf r}')\,,
\ee
where we used the fact that $\mathcal{G}_{VV}^{(0)}$ does not contribute to the opening of the wormhole. 
Using (\ref{eq-h}) and performing the integral in $V_1$, the expectation value of the stress-energy tensor can be written as 
\be \label{eq-stress1}
\langle T_{VV} \rangle = \frac{h}{4\pi^3} \frac{D_c^4 G^{xx}(z_h)}{T}  \int \frac{d^{d-1}k}{(2\pi)^{d-1}} \frac{1}{  \sin (\pi D_k) }  \frac{k^6}{\left[(V/V_0-1)(1+V V_0)\right]^{D_k+1}} \,.
\ee
By writing $k$ in spherical coordinates,
the expectation value of the stress-energy tensor can then be written as
\begin{align} \label{eq-Tvv-num1}
\langle T_{VV} \rangle = \frac{h}{2^d \pi^{\frac{d+5}{2}}\Gamma\left( \frac{d-1}{2}\right)}& \frac{ D_c^4 G^{xx}(z_h)}{T} \frac{1}{P(V,V_0)} \int_{0}^{k_\text{max}} dk \frac{ k^{d+4} }{  \sin (\pi D_k) P(V,V_0)^{D_k} } \,, \\
&P(V,V_0)=\left(\frac{V}{V_0}-1\right)(1+V V_0)\,,
\end{align}
and $k_\text{max}$ is a momentum cutoff. Since we derived the bulk-to-boundary propagators using a hydrodynamic approximation, in which $\frac{k}{2\pi T} \ll 1$, it is natural to set $k_\text{max} \sim 2 \pi T$. In this work, however, we choose to introduce a double trace deformation that only involves low energy modes, and we write the momentum cutoff as $k_\text{max} =  2 \pi T \sqrt{f}$, where the parameter $f$ controls the maximum energy of our deformation. This is conceptually similar to what was done in \cite{Maldacena:2017axo, Gao:2018yzk}, in which the authors consider a signal that only contains low energy modes by changing the corresponding wave functions as $\psi(p) \rightarrow \psi(p) e^{-p^2/\sigma^2}$. Here we do a similar procedure for the deformation, instead of the signal, and introduce a hard cutoff, instead of a smooth (Gaussian) cutoff. Our motivation to consider such type of deformation is because we are interested in the connection between traversability and diffusion, and diffusion only takes place for low energy modes. Before evaluating the integral in (\ref{eq-Tvv-num}), let us first introduce a new variable, $u = D_k = \frac{D_c}{2\pi T}k^2$, and factor out the temperature dependence of $h$, $D_c$ and $G^{xx}(z_h)$ by writing them as follows $h= \tilde{h}T^{2-d}$, $D_c=\mathcal{D}_c/T$ and $G^{xx}=\tilde{G}^{xx}/T^2$. In terms of these new variables, the expectation value of the stress-energy tensor can be written as
\be \label{eq-Tvv-num}
\langle T_{VV} \rangle = \mathcal{N}(\tilde{h},z_h,d,\mathcal{D}_c)  \frac{1}{P(V,V_0)}\int_{0}^{u_\text{max}} du \frac{u^{\frac{d+3}{2}}}{ \sin(\pi u) P(V,V_0)^u}\,, \\
\ee
where the overall factor
\be \label{eq-N}
\mathcal{N}(\tilde{h},z_h,d,\mathcal{D}_c)= \frac{\tilde{h}\, \tilde{G}^{xx}(z_h) \mathcal{D}_c^{\frac{3-d}{2}}}{2^{\frac{d-3}{2}} \Gamma\left( \frac{d-1}{2}\right)}
\ee
contains information about the geometry, while the remaining part only depends on $u_\text{max}$. Note that $k_\text{max} =   2\pi T \sqrt{f}$ implies $u_\text{max} = 2 \pi \mathcal{D}_c f $, which is typically a $\mathcal{O}(1)$ number\footnote{To prevent instances where the integral with respect to $u$ in (\ref{eq-Tvv-num}) encounters the pole at $u=1$, we establish an UV cutoff by ensuring that $u_\text{max}$ remains below 1. In scenarios where $u_\text{max}$ exceeds 1, one could potentially compute the stress-energy tensor by numerically extracting the principal value of the integral in (\ref{eq-Tvv-num}). However, addressing these intricacies is beyond the scope of our current investigation.}. Fig.~\ref{fig-Tvv3} shows the behavior of $ \mathcal{N}^{-1}\langle T_{VV} \rangle$ as a function of $V$ for several values of $V_0$ with $d=4$ and $u_\text{max}=1/2$. This figure shows that the stress-energy tensor diverges at the insertion time $V=V_0$, and quickly decreases to zero for larger values of $V$. Similar behavior was also observed in \cite{Gao:2016bin}.

\subsection{Averaged null energy}
\begin{figure}[H]
    \centering
    \includegraphics[width=0.55\textwidth]{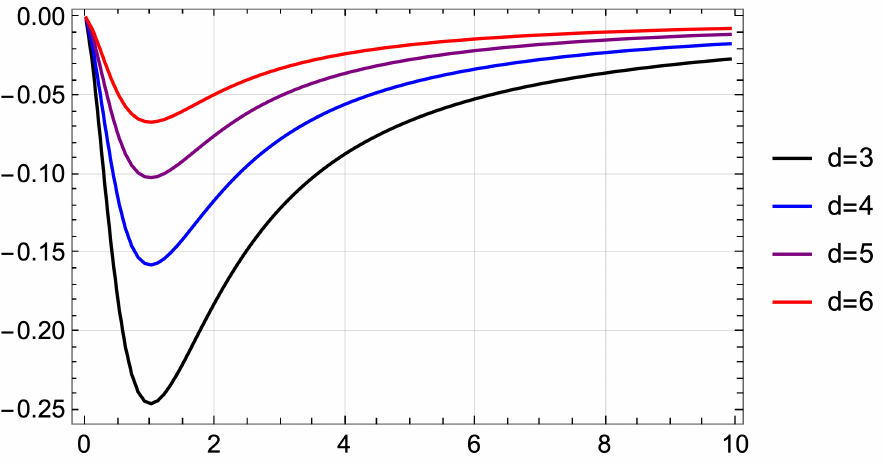}
    \put(-130,-10){$V_0$}
    \put(-257,35){\rotatebox{90}{$ \mathcal{N}^{-1} \int \langle T_{VV} \rangle dV $}}    \caption{ averaged null energy  versus $V_0$ in asymptotically AdS$_{d+1}$ spacetimes for several values of $d$. Here we fix $u_\text{max}=1/2$ and $\tilde{h}=-1$.}
    \label{fig-DeltaU}
\end{figure}

In this subsection, we compute the averaged null energy in the presence of double trace deformation involving $U(1)$ conserved current operators. By integrating \eqref{eq-Tvv-num} along complete achronal null geodesics, the averaged null energy can be computed as
\be \label{eq-ANEC-num}
\mathcal{A}=\int_{V_0}^{\infty} dV \langle T_{VV} \rangle = \mathcal{N}(\tilde{h},z_h,d,\mathcal{D}_c) \int_{V_0}^{\infty} dV \frac{1}{P(V,V_0)}\int_{0}^{u_\text{max}} du \frac{u^{\frac{d+3}{2}}}{ \sin(\pi u) P(V,V_0)^u}\,,
\ee
where $\mathcal{N}(\tilde{h},z_h,d,\mathcal{D}_c)$ is given by (\ref{eq-N}). The above integrals can be computed numerically. In Fig.~\ref{fig-DeltaU}, we plot $ \mathcal{N}^{-1} \int T_{VV} dV$ versus $V_0$ for several spacetime dimensions using (\ref{eq-ANEC-num}). The averaged null energy becomes less negative as we increase $d$, suggesting that it is more difficult to open the wormhole in higher dimensional setups, in accordance with previous results in the literature \cite{Ahn:2020csv}.

The relation between the averaged null energy and the wormhole opening $\Delta U$ reads (see Appendix \ref{sec-app0})
\be
\Delta U =-\frac{1}{G_{UV}(0)} \frac{16 \pi G_N }{d(d-1)} \int_{V_0}^{\infty} \langle T_{VV} \rangle dV\,.
\ee
Using (\ref{eq-ANEC-num}), we write the wormhole opening as follows
\be \label{eq-ANEC-final1}
\Delta U =-\frac{\tilde{h}}{G_{UV}(0)} \frac{16 \pi G_N }{d(d-1)} \int_{V_0}^{\infty} \frac{dV}{P(V,V_0)} \, \frac{1}{\mathcal{D}_c^{\frac{d-3}{2}}}\int_{0}^{2\pi f \mathcal{D}_c} du\, \mathcal{F}(u,V,V_0,d)
\ee
where
\be \label{eq-ANEC-final2}
\mathcal{F}(u,V,V_0,d) =  \frac{ \tilde{G}^{xx}(z_h) }{2^{\frac{d-3}{2}} \Gamma\left( \frac{d-1}{2}\right)} \frac{u^{\frac{d+3}{2}}}{ \sin(\pi u) P(V,V_0)^u}\,.
\ee
 Here we write the UV cutoff as $k_\text{max}=2 \pi T \sqrt{f} $, which leads to $u_\text{max}= 2\pi \mathcal{D}_c f$. The factor $f$ controls the size of the UV cutoff.

\subsubsection{Application for linear axion models} \label{sec-axionic}
In this subsection, we would like to investigate how the averaged null energy depends on the charge diffusion constant. In order to do that, we consider a linear axion model where the charge diffusion constant depends on a parameter that controls the momentum relaxation of the dual field theory, and becomes arbitrarily small as one increases the momentum relaxation parameter.

More specifically, we consider a simplified version of the linear axion model proposed in \cite{Andrade:2013gsa}, with action of the form
\be
S= \frac{1}{16 \pi G_N}\int d^{d+1}x \sqrt{-g} \left[R-\frac{d(d-1)}{L^2} -\frac{1}{2}\sum_{i}^{d-1}(\partial \psi_I)^2\right]\,,
\ee
where $L$ is the AdS length scale. For convenience, we set $L=1$.
We consider the background solution found in \cite{Andrade:2013gsa}, which takes the form
\be \label{eq-solaxion}
ds^2=-f(R)dt^2+\frac{dR^2}{f(R)}+R^2\delta_{ab}dx^a dx^b\,,\,\,\,\,\,\, \psi_I =\alpha_{Ia}\, x^a
\ee
where the boundary is located at $R \rightarrow \infty$, and the black hole horizon is located at $R=R_0$. Here, the index $a$ runs from $1$ to $d-1$, labeling the $d-1$ spatial directions, $I$ is an internal index labeling $d-1$ scalar fields, and $\alpha_{Ia}$ are real arbitrary constants. The emblackening factor is given by 
\be
f(r) =R^2-\frac{\alpha^2}{2(d-2)}-\frac{R_0^d}{R^{d-2}}\left(1-\frac{1}{2(d-2)}\frac{\alpha^2}{R_0^2} \right)\,.
\ee
where $\alpha$ is a parameter that controls the momentum relaxation of the system, and it is given by
\be
\alpha^2 =\frac{1}{d-1}\sum_{a=1}^{d-1} \Vec{\alpha}_a \cdot \Vec{\alpha}_a\,,
\ee
where $\Vec{\alpha}_a \cdot \Vec{\alpha}_b =\sum_{I=1}^{d-1} \alpha_{Ia} \alpha_{Ib}$, and we made the assumption that $\Vec{\alpha}_a \cdot \Vec{\alpha}_b = \alpha^2 \, \delta_{ab}$ for all values of $a$ and $b$.
The black hole temperature is given by
\be \label{eq-Temp}
T=\frac{f'(R_0)}{4\pi}=\frac{1}{4\pi} \left(d R_0-\frac{\alpha^2}{2R_0^2} \right)\,.
\ee
Since there is no background gauge field in the above solution, the charge diffusion constant associated to fluctuations of a probe gauge field can be computed using Eq.~(\ref{eq-diffusionDc}) derived in Appendix \ref{app-new}:
\be \label{eq-diffusionDc2}
D_c=\frac{\sqrt{-G}}{G_{xx}\sqrt{G_{tt}G_{zz}}}\Big|_{z_h} \int_{0}^{z_h} dz \frac{G_{tt}G_{zz}}{\sqrt{-G}}=\frac{2 d}{(d-2) \left(4 \pi  T+ \sqrt{2d \alpha ^2 +16 \pi ^2 T^2}\right)}\,.
\ee
The dimensionless charge diffusion constant takes the form
\be
\mathcal{D}_c=D_c T= \frac{2 d}{(d-2) \left(4 \pi  + \sqrt{2d \frac{\alpha^2}{T^2} +16 \pi ^2 }\right)}\,.
\ee
For $\alpha=0$, it takes the value $\mathcal{D}_c=\frac{d}{d-2} \frac{1}{4\pi}$, and it becomes arbitrarily small in the zero temperature limit, in which the ratio $\alpha/T$ becomes arbitrarily large.\footnote{Note that to have $T \geq 0$ in (\ref{eq-Temp}), the momentum relaxation parameter $\alpha$ cannot take arbitrarily large values. Therefore, to have an arbitrarily large value of the ratio $\alpha/T$ we need to take the small temperature limit. }

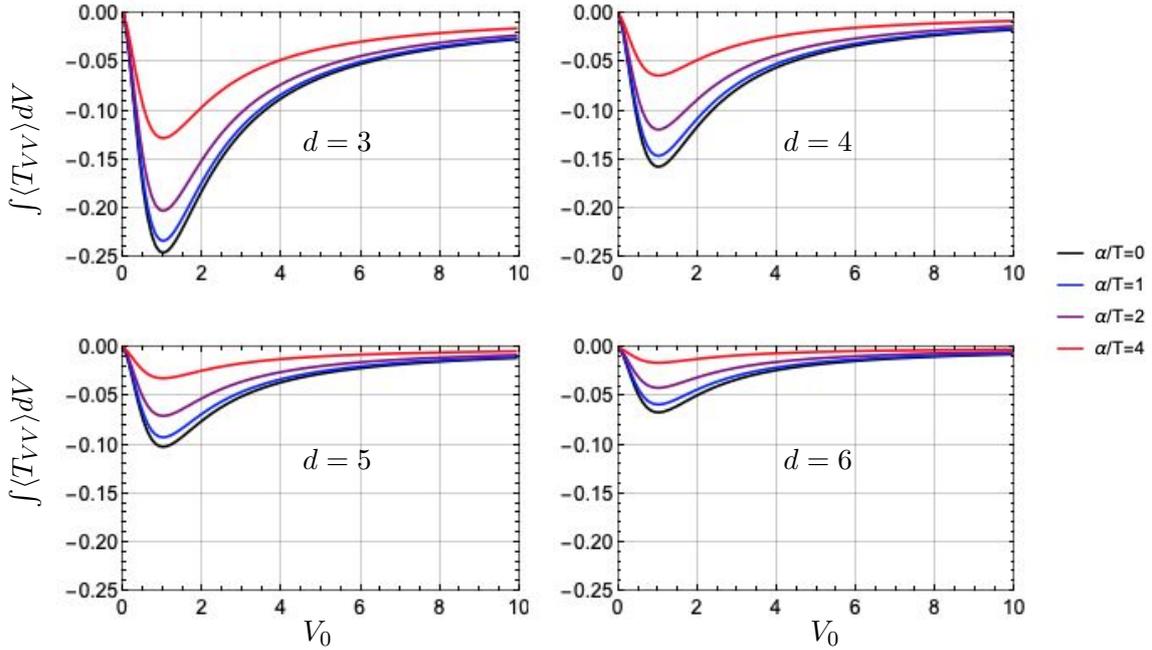
\begin{figure}[h!]
    \centering
    \includegraphics[width=1\textwidth]{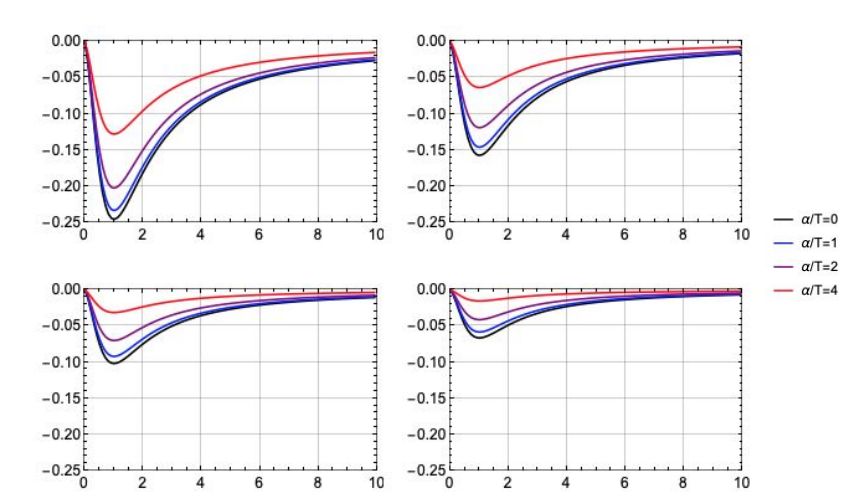}
    \put(-430,45){\rotatebox{90}{$  \int \langle T_{VV} \rangle dV $}}
     \put(-430,155){\rotatebox{90}{$  \int \langle T_{VV} \rangle dV $}}
     \put(-130,-5){$V_0$}
     \put(-320,-5){$V_0$}
    \put(-320,180){$d=3$}
    \put(-140,180){$d=4$}
    \put(-320,60){$d=5$}
    \put(-140,60){$d=6$}
    \caption{$ \int T_{VV} dV$ as a function of $V_0$ for several values of the momentum relaxation parameter $\alpha/T$ and $d$ for the asymptotically AdS$_{d+1}$ background solution in (\ref{eq-solaxion}). Here we fix $u_\text{max}=1/2$ for $\alpha=0$ by setting $f=(d-2)/d$.}
    \label{fig:ANEvsALPHA}
    
\end{figure}

We would like to understand how the presence of momentum relaxation affects the traversability properties of the wormhole. In our hydrodynamic limit, our result is almost universal, and the dependence on the system comes basically from the dimensionless charge diffusion constant, which naively appears only as an overall factor of $\mathcal{D}_c^{\frac{3-d}{2}}$. In the regime of strong momentum relaxation, $\mathcal{D}_c$ becomes arbitrarily small, and the overall factor of $\mathcal{D}_c^{\frac{3-d}{2}}$ becomes arbitrarily large, causing $\Delta U$ to diverge. However, once we introduce an UV cutoff of the form $k_\text{max}= 2\pi T \sqrt{f}$, this implies that the wormhole opening takes the schematic form
\be \label{eq-DeltaUschematic}
\Delta U \sim \tilde{h}\, G_N\,\int_{V_0}^{\infty} \frac{dV}{P(V,V_0)} \frac{1}{\mathcal{D}_c^{\frac{d-3}{2}}}\int_{0}^{2\pi f \mathcal{D}_c} du \, \mathcal{F}(u,V,V_0,d)\,,
\ee
and the limit of strong momentum relaxation (in which $\mathcal{D}_c \rightarrow 0$) does not lead to any divergence of $\Delta U$, due to the fact that the upper limit of integration is also proportional to the dimensionless charge diffusion constant, i.e., $u_\text{max} = 2\pi  \mathcal{D}_c f$. In Figure~\ref{fig:ANEvsALPHA} we show the result for the averaged energy condition for several spacetimes dimensions and for several values of $\alpha/T$, which is the parameter  that controls the momentum relaxation in the system. From this figure, we can see that the wormhole opening decreases as we increase the momentum relaxation parameter, and approaches zero as $\alpha/T \gg 1$. That suggests that, at least in our hydrodynamic approximation, the presence of momentum relaxation does not favor traversability. In fact, a small diffusion constant leads to a very small wormhole opening\footnote{In previous versions of this paper, we speculated that a small diffusion constant could favor traversability. The more precise calculation of the current version of this paper actually shows otherwise.}. 

\subsection*{Fitting the data}
\begin{figure}[h!]
    \centering
    \includegraphics[width=1\textwidth]{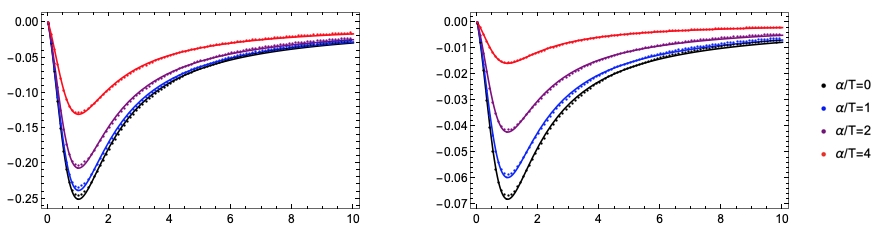}
    \put(-445,35){\rotatebox{90}{$  \int \langle T_{VV} \rangle dV $}}
    \put(-235,35){\rotatebox{90}{$  \int \langle T_{VV} \rangle dV $}}
    \put(-370,0){$V_0$}
    \put(-350,50){$d=3$}
    \put(-145,0){$V_0$}
    \put(-125,50){$d=6$}
    \caption{$ \int T_{VV} dV$ as a function of $V_0$ for several values of the momentum relaxation parameter $\alpha/T$. The dots represent numerical results obtained with Eq.~(\ref{eq-ANEC-num}) and the continuous curves represent the fitted curves of the form (\ref{eq-fitted})}
    \label{fig:fittedData}
\end{figure}

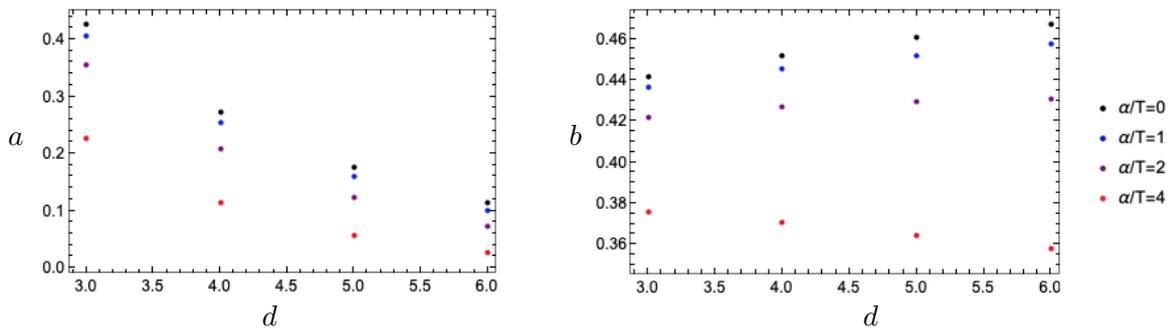
\begin{figure}[h!]
    \centering
    \includegraphics[width=1\textwidth]{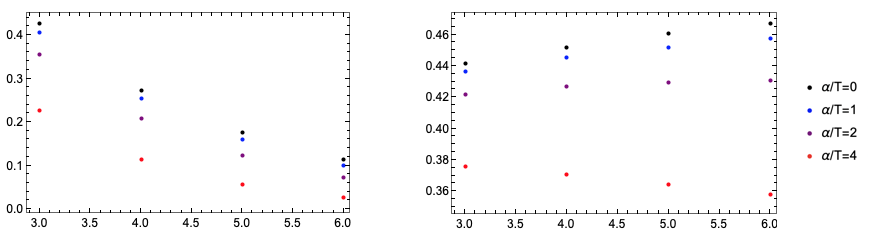}
    \put(-440,65){$a$}
    \put(-230,65){$b$}
    \put(-120,-3){$d$}
    \put(-345,-3){$d$}
    \caption{Fitted parameters $a$ and $b$ versus $d$ for several values of the momentum relaxation parameter $\alpha/T$. The fit was done in the interval $V_0 \in [0,10]$.}
    \label{fig:fittedParameters}
\end{figure}

In this subsection, we show that our numerical results for the averaged null energy can be described reasonably well by the following function:
\be \label{eq-fitted}
\mathcal{A}(V_0) = \int_{V_0}^{\infty} dV \langle T_{VV} \rangle = -\frac{  V_0 }{V_0^2+1} \frac{a}{\left[ \log \left(V_0+\frac{1}{V_0}\right)\right]^{b}}
\ee
where the fitting parameters $a$ and $b$ are functions of $d$ and $\alpha/T$.
Figure~\ref{fig:fittedData} shows our numerical results for the averaged null energy, obtained with Eq.~(\ref{eq-ANEC-num}) as well as the fitted function of the form (\ref{eq-fitted}). Figure~\ref{fig:fittedParameters} shows the results for the fitted parameters $a$ and $b$ for several values of $d$ and $\alpha/T$. The fit was done in the interval $V_0 \in [0,10]$. The parameter $a$ decreases as we increase the dimensionality of spacetime or the momentum relaxation parameter. The parameter $b$, however, does not change very much as we increase $d$, and decreases as we increase $\alpha/T$.

\subsection{Bound on information transfer } \label{sec-bound}
A traversable wormhole allows us to send a signal from one side of the geometry to the other side. In this section, we derive a bound on the amount of information that can be transferred through wormholes opened by conserved current operators and discuss what happens to this bound in the small frequency limit.
The basic idea is that the backreaction of each particle that we send through the wormhole has the effect of reducing $|\Delta U|$, i.e., it closes the wormhole a bit. Therefore, if we try to send a signal with too many particles, the corresponding backreaction will eventually reduce $|\Delta U|$ to zero, closing the wormhole completely. 

Consider a message with $N_\text{bits}$ particles, each one with momentum $p_U^\text{each}$, such that the total momentum of the signal is $p_U^\text{tot}=N_\text{bits}\, p_U^\text{each}$. Using the uncertainty principle and requiring that the signal `fits' in the opening of the wormhole. The uncertainty principle implies that
\be \label{eq-unc}
p_U^\text{each} \Delta U_\text{each} \gtrsim 1\,.
\ee
Let us assume that the double trace deformation open the wormhole by an amount $\Delta U$. Requiring that each particle's wave function fits in the wormhole opening, we find that
\be \label{eq-fit}
\Delta U_\text{each} \leq |\Delta U|.
\ee
Combining (\ref{eq-unc}) and (\ref{eq-fit}), we find
\be \label{eq-aux}
\frac{1}{p_U^\text{each}} \lesssim \Delta U_\text{each} \leq |\Delta U|\,.
\ee
Using (\ref{eq-aux}), and the fact that $N_\text{bits} = \frac{p_U^\text{tot}}{p_U^\text{each}}$, we find
\be \label{eq-bound1}
N_\text{bits} \lesssim p_U^\text{tot} |\Delta U|\,.
\ee
Finally, we require that the backreaction of the signal is small. To impose this condition, we need to model the stress-energy tensor of the signal. For a very early perturbation, the stress-energy of the signal takes a simple form, which can be written as $T_{UU} \sim \frac{p^\text{tot}_U}{\sqrt{G_{xx}^{d-1}(z_h)}} \delta(U)$. The probe approximation requires $G_N \frac{p^\text{tot}_U}{\sqrt{G_{xx}^{d-1}(z_h)}} \ll 1$, and it breaks down roughly at
\be \label{eq-probe}
p^\text{tot}_U\lesssim \frac{\sqrt{G_{xx}^{d-1}(z_h)}}{G_N}\,.
\ee
Finally, plugging (\ref{eq-probe}) and (\ref{eq-DeltaUschematic}) into (\ref{eq-bound1}), and we obtain
\be \label{eq-bound2}
N_\text{bits} \lesssim  \tilde{h} \, \left( \sqrt{G_{xx}(z_h)} \right)^{d-1} \int_{V_0}^{\infty} \frac{dV}{P(V,V_0)} \frac{1}{\mathcal{D}_c^{\frac{d-3}{2}}}\int_{0}^{2\pi f \mathcal{D}_c} du \, \mathcal{F}(u,V,V_0,d) \,.
\ee
The fact that the bound is proportional to $\left( \sqrt{G_{xx}(z_h)} \right)^{d-1}$ is consistent with the analysis of \cite{Freivogel_2020} for higher dimensional traversable wormholes. The functional dependence of the result on $\mathcal{D}_c$ implies that $N_\text{bits} \rightarrow 0$ as $\mathcal{D}_c \rightarrow 0$.  From the point of view of the boundary theory, that suggests that the presence of momentum relaxation makes the teleportation protocol less efficient. It would be interesting to check if this also happens in many-body quantum teleportation protocols in which the classical communication involves the measurement of conserved current operators.


\section{Discussion } \label{sec-disc}

We constructed a traversable wormhole by coupling the two asymptotic boundaries of a general $(d+1)$-dimensional black brane with $U(1)$ conserved current operators. The non-local coupling introduces a quantum correction to the expectation value of the stress-energy tensor that violates ANEC, rendering the wormhole traversable.

\begin{figure}[H]
    \centering
    \includegraphics[width=0.5\textwidth]{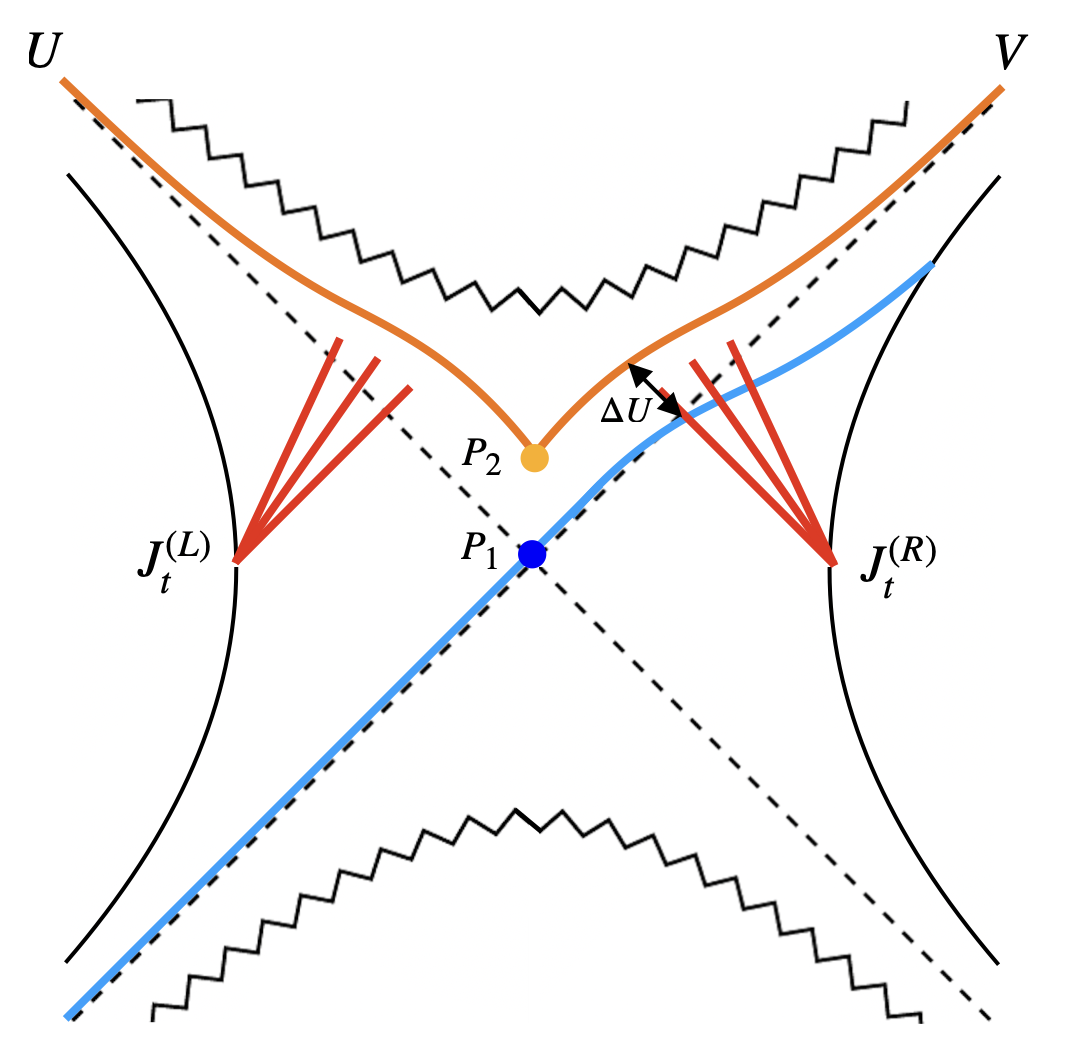}
    \caption{The boundary double trace deformation with $U(1)$ conserved current operators corresponds to gauge field fluctuations in the bulk. One can view the red lines as photons that are delocalized with different longitudinal energies, and it indicates the diffusive behavior of the bulk gauge fields. The right and left past (future) horizons intersect at the point $P_1 (P_2)$. The size of the wormhole opening $\Delta U$ can be obtained by integrating along an achronal null geodesic, which is represented by the blue line.}
    \label{fig-diffusive}
\end{figure}

The double trace deformation involving $U(1)$ conserved current operators is dual to a gauge field fluctuation in the bulk. In the limit where the frequency is small, the gauge field in the bulk displays a diffusive behavior. We found that the diffusive properties of the gauge field affect the behavior of the wormhole opening $\Delta U$ (see Fig.~\ref{fig-diffusive}), causing it to have a power-law behavior as a function of the insertion time $t_0$. 
In fact, by using that $V_0 = e^{2\pi T t_0}$, and taking the limit where $2\pi T t_0 \gg 1$ in (\ref{eq-fitted}), we can write the averaged null energy as
\be
\mathcal{A}(t_0) \sim \frac{ e^{-2\pi T t_0} }{t_0^{b}}\,.
\ee
The power law part is reminiscent of the power law behavior observed in two-point functions \cite{Kovtun_2003, Caron-Huot:2009kyg} and out-of-time-order correlators involving conserved current operators \cite{Cheng:2021mop}, and it is related to the diffusive behavior of the bulk gauge field. The exponential part also takes place in the large time limit when the double trace involves scalar operators. For scalar operators of dimension $\Delta$, one finds \cite{Freivogel_2020,Ahn:2020csv}
\be
\mathcal{A}_\text{scalar} \sim \left( \frac{V_0}{1+V_0^2}\right)^{2\Delta+1}\,.
\ee
In the limit $t_0 \gg 1/T$, one obtains $\mathcal{A}_\text{scalar} \sim e^{-2\pi T (2\Delta+1) t_0}$. Therefore, while the exponential decrease with $t_0$ is present in both cases, the power law part only appears in the case involving conserved current operators.

We studied how the wormhole opening depends on the charge diffusion constant. In order to do that, we consider two-sided axionic black hole solutions (see Sec.~\ref{sec-axionic}) and study the behavior of the averaged null energy as a function of the momentum relaxation parameter. In these backgrounds, the charge diffusion constant becomes arbitrarily small as one increases the momentum relaxation parameter $\alpha/T$. We computed the averaged null energy in the same limit and checked that it becomes arbitrarily small as $\alpha/T \gg 1$. See Fig.~\ref{fig:ANEvsALPHA}. This suggests that the presence of momentum relaxation makes it harder to open the wormholes using $U(1)$ conserved current operators. 

 To compute the charge diffusion constant, we introduced a bulk gauge field in the probe approximation. If the bulk gauge field backreacts in the geometry, then our formula for the charge diffusion constant Eq.~(\ref{eq-diffusionDc}) is not valid. It would be interesting to generalize our results for systems with non-zero chemical potential.  Another interesting future direction would be to consider double trace deformations involving other types of conserved currents and study how the wormhole opening depends on the corresponding transport coefficients.

\acknowledgments

We would like to thank Mitsuhiro Nishida, Kyung Kiu Kim, Hyun Seok Yang, Kyoung-Bum Huh and Hyun-Sik Jeong for helpful discussions. This work was supported by Basic Science Research Program through the National Research Foundation of Korea(NRF) funded by the Ministry of Science, ICT \& Future
Planning (NRF- 2021R1A2C1006791), the GIST Research Institute(GRI) and  the AI-based GIST Research Scientist Project grant funded by the GIST in 2022. 
V. Jahnke was supported by Basic Science Research Program through the National Research Foundation of Korea(NRF) funded by the Ministry of Education(NRF-2020R1I1A1A01073135). B. Ahn was also supported by Basic Science Research Program through the National Research Foundation of Korea funded by the Ministry of Education (NRF-2020R1A6A3A01095962, NRF-2022R1I1A1A01064342).


\appendix

\section{ANEC violation} \label{sec-app0}

In this appendix, we review how the wormhole opening is related to the violation of ANEC. We consider a background metric  $G_{\mu \nu}$ that satisfies (\ref{eq-eom}) with vanishing right hand side. Assuming the double trace deformation introduces a non-zero expectation value for the stress energy tensor $\langle T_{\mu \nu} \rangle \sim \mathcal{O}(\epsilon)$, we consider fluctuations $G_{\mu \nu} \rightarrow G_{\mu \nu} + h_{\mu \nu}$, with\\ $h_{\mu \nu} \sim \mathcal{O}(\epsilon)$. Using Kruskal coordinates, we find that at $U=0$,

\be
\frac{1}{2} \left(-\Lambda\, h_{VV}+\delta^{ij}\frac{ \partial_V h_{ij}+ V   \partial_V^2 h_{ij} }{V z_h^2 } \right) = 8 \pi G_N \langle T_{VV}\rangle\,.
\ee
Integrating both sides with respect to $V$ we find
\be
8 \pi G_N \int \langle T_{VV} \rangle dV = -\Lambda \int h_{VV} dV\,.
\ee
The wormhole opening can then be computed as 
\be
\Delta U = -\frac{1}{G_{UV}(0)} \int  h_{VV} dV = -\frac{1}{G_{UV}(0)} \frac{16 \pi G_N}{d(d-1)} \int \langle T_{VV} \rangle dV\,.
\ee
where in the last equality we used that $\Lambda = -\frac{d(d-1)}{2L^2}$. The metric component in the $UV$ direction, denoted as $G_{UV}$, is expressed as $G_{UV}=-\frac{\beta^2}{4\pi^2}\frac{G_{tt}}{UV}$, which can be determined once the geometry is specified. Near the horizon, the tortoise coordinate is given by $r=\frac{\beta}{4\pi}\log\frac{z-z_h}{k_1}$, where $k_1$ is a constant that depends on the geometry, allowing us to evaluate the product $UV$. In the exterior regions, $UV=-e^{\frac{4\pi}{\beta}r}=-\frac{z-z_h}{k_1}$. By using the near-horizon expression for $G_{tt}$, namely $G_{tt}=c_0(z-z_h)$, we obtain $G_{UV}(0) =\frac{\beta^2}{4\pi^2} c_0 k_1$.

\section{Vector field fluctuation on a generic black brane background } \label{app-new}
In this appendix, we review how to solve the Maxwell-Einstein equation of motion
\be \label{eq-gauge eom0}
\partial_\mu (\sqrt{-g} F^{\mu \nu}) =0\,,
\ee
for a generic black brane background with line element of the form
\be \label{eq-metricapp}
ds^2 = g_{tt} dt^2+g_{rr}dr^2+g_{xx} \delta_{ij}dx^idx^j
\ee
in the hydrodynamic limit. In (\ref{eq-metricapp}), $r$ is tortoise coordinate, defined in (\ref{eq-tortoise}) and $g_{\mu \nu}$ denote the metric components in tortoise coordinates $(t,r,x^i)$. In terms of the general line element (\ref{eq-background}), these metric components are written as $g_{tt}=-g_{rr}=-G_{tt}$, and $g_{xx}=G_{xx}$.
For simplicity, we follow~\cite{Cheng:2021mop} and we choose a configuration in which $A_i=0$. This corresponds to analyzing the longitudinal sector of the gauge field fluctuations, which are known to be controlled by a diffusion pole~\cite{Iqbal:2008by, Starinets:2008fb}.

We first decompose the gauge field in Fourier modes
\be \label{eq-Fourier}
A_{\mu} =\int \frac{d\omega d^{d-1} k}{(2\pi)^{d}} \, A_{\mu} (r,\omega,k) e^{-i \omega t} e^{i k \cdot x}\,,
\ee
and assume the momentum to be parallel to the $x$ axis. With these assumptions, Eq.~(\ref{eq-gauge eom0}) for $\nu = r$ becomes 
\be \label{eq-eom-r}
-i \omega (\sqrt{-g}g^{rr}g^{tt}(-i \omega A_r - \partial_r A_t))- k^2 (\sqrt{-g}g^{xx}g^{tt} A_r)=0\,.
\ee
Solving (\ref{eq-eom-r}) for $A_r$, we find
\be \label{eq-Ar}
A_r =\frac{i \omega g^{tt} \partial_r A_t}{\omega^2 g^{tt}+k^2g^{xx}}\,.
\ee
Now we turn our attention to Eq.~(\ref{eq-gauge eom0}) for $\nu =t$:
\be \label{eq-eom-t}
\partial_r (\sqrt{-g}g^{rr}g^{tt}(\partial_r A_t + i \omega A_r))-k^2 (\sqrt{-g}g^{xx}g^{tt} A_t)=0\,,
\ee
Substituting (\ref{eq-Ar}) into (\ref{eq-eom-t}), we find an equation for $A_t$:
\be \label{eq-onlyAt}
A_t''(r)+ A_t'(r)\, \partial_r \log \left[  \frac{ \sqrt{-g}g^{rr}g^{tt} g^{xx}}{\omega^2 g^{tt}+k^2 g^{xx}} \right]  - g_{rr}(\omega^2 g^{tt}+k^2 g^{xx}) A_t(r) =0\,.
\ee
The above equation has two independent solutions, which near the horizon ($r \rightarrow \infty$) take the form $e^{\pm i \omega r}$, corresponding to out-going and in-falling boundary conditions. We focus on the solution that satisfies in-falling boundary conditions at the horizon, and look for a solution of the form
\be
A_t(r)= e^{-i \omega r} F(r)
\ee
With this ansatz, Eq.~(\ref{eq-onlyAt}) becomes
\be
F''-2 i \omega F'+\partial_r\log \left[  \frac{ \sqrt{-g}g^{rr}g^{tt} g^{xx}}{\omega^2 g^{tt}+k^2 g^{xx}} \right]\left(F'-i \omega F\right)-\left[(1+g_{rr}g^{tt})\omega^2+g_{rr}g^{xx}k^2 \right] F=0\,,
\ee
To find a solution in the hydrodynamic limit, we set $(\omega, k) \rightarrow \lambda (\omega, k)$ and take $\lambda \ll 1$. We then expand the function $F$ as follows
\be
F= F_0 + \lambda F_1 + \cdots
\ee
and solve the equations of motion at each order in the parameter $\lambda$. At zero order in $\lambda$, we have the following equation
\be
F_0''+\partial_r\log \left[  \frac{ \sqrt{-g}g^{rr}g^{tt} g^{xx}}{\omega^2 g^{tt}+k^2 g^{xx}} \right]F_0'=0\,,
\ee
whose solution reads
\be
F_0=C_0+C_1 \int^{r} dr'   \frac{\omega^2 g^{tt}+k^2 g^{xx}}{ \sqrt{-g}g^{rr}g^{tt} g^{xx}} \,.
\ee
To find a regular solution it is convenient to write $F_0$ in terms of the coordinates $(t,x^i,z)$ defined in Sec.~\ref{sec-gravity}. In these coordinates, we find
\be
F_0=C_0+C_1 \int^{z}dz' \sqrt{\frac{G_{zz}}{G_{tt}}}G_{xx}^{\frac{3-d}{2}} \left(-\omega^2+\frac{G_{tt}}{G_{xx}}k^2 \right)\,.
\ee
In the near horizon region, the solution takes the approximate form
\be
F_0=C_0-C_1 \frac{\omega^2 \beta}{4\pi}G_{xx}(z_h)^{\frac{3-d}{2}} \log \left(z-z_h\right)\,,
\ee
For this solution to be regular at the horizon we need to set $C_1=0$. At linear order in $\lambda$, we find
\be
F_1''-2 i \omega F_0'-\partial_r \log \left[ \frac{\omega^2 g^{tt}+k^2 g^{xx}}{ \sqrt{-g}g^{rr}g^{tt} g^{xx}} \right](F_1'-i\omega F_0)=0\,.
\ee
Using that $F_0=C_0$, we find
\be
F_1 = \int^{r}dr'\, \left[ i \omega C_0+C_2\left( \frac{\omega^2 g^{tt}+k^2 g^{xx}}{ \sqrt{-g}g^{rr}g^{tt} g^{xx}}\right) \right]+C_3\,.
\ee
In terms of the coordinates $(t,x^i,z)$, the solution reads
\be
F_1= - \int^{z} dz' \, \sqrt{\frac{G_{zz}}{G_{tt}}} \left[i \omega C_0+C_2\left(\omega^2-\frac{G_{tt}}{G_{xx}}k^2 \right)G_{xx}^{\frac{3-d}{2}} \right]+C_3\,.
\ee
In the near horizon region, the solution takes the approximate form
\be
F_1=-\frac{\beta}{4\pi} \left( i \omega C_0+ C_2 \omega^2 G_{xx}(z_h)^{\frac{3-d}{2}} \right) \log \left(z-z_h \right)+C_3\,.
\ee
Requiring regularity at the horizon fixes the integration constant $C_2$ as
\be
C_2=-\frac{i C_0 }{\omega} (G_{xx}(z_h))^{\frac{d-3}{2}}\,.
\ee
Then, we can use $C_3$ to write the solution in terms of a definite integral
\be
F_1=  i \omega C_0 \int_{z}^{z_h} dz' \, \sqrt{\frac{G_{zz}(z)}{G_{tt}(z)}} \left[1-\frac{G_{xx}(z_h)^{\frac{d-3}{2}}}{G_{xx}(z)^{\frac{d-3}{2}}}\left(1-\frac{G_{tt}(z)}{G_{xx}(z)}\frac{k^2}{\omega^2} \right) \right],.
\ee
where we wrote the argument of the metric functions explicitly to avoid confusion.
To first order in $\lambda$, the gauge field can be written as
\be
A_t(z)=C_0\, e^{-i \omega r(z)} \left( 1+ i \omega \int_{z}^{z_h} dz \, \sqrt{\frac{G_{zz}(z)}{G_{tt}(z)}}  \left[1-\frac{G_{xx}(z_h)^{\frac{d-3}{2}}}{G_{xx}(z)^{\frac{d-3}{2}}}\left(1-\frac{G_{tt}(z)}{G_{xx}(z)}\frac{k^2}{\omega^2} \right) \right]  \right)\,.
\ee
The constant $C_0$ can be fixed by requiring $A_t(z \rightarrow 0)=1$. To do that, it is convenient to write the solution as
\be
A_t(z)=C_0\, e^{-i \omega r(z)} \left( 1+ i \omega H_1(z)+\frac{i k^2}{\omega} H_2(z) \right)\,.
\ee
where
\be
H_1(z)=\int_{z}^{z_h} dz \, \sqrt{\frac{G_{zz}(z)}{G_{tt}(z)}} \left(1- \frac{G_{xx}(z_h)^{\frac{d-3}{2}}}{G_{xx}(z)^{\frac{d-3}{2}}}\right)\,,\,\,\,\,H_2(z)=\int_{z}^{z_h} dz \,  \sqrt{G_{zz}(z) G_{tt}(z)}\frac{G_{xx}(z_h)^{\frac{d-3}{2}}}{G_{xx}(z)^{\frac{d-1}{2}}}\,.
\ee
The constant $C_0$ can then be obtained as
\be
C_0=\frac{1}{1+i \omega H_1(0)+\frac{i k^2}{\omega}H_2(0)}\,.
\ee
The normalized gauge field then becomes
\be
A_t(z)=C_0\, e^{-i \omega r(z)} \frac{\left( 1+ i \omega H_1(z)+\frac{i k^2}{\omega} H_2(z) \right)}{\left( 1+ i \omega H_1(0)+\frac{i k^2}{\omega} H_2(0) \right)}\,.
\ee
The above expression has a pole which, at leading order in a hydrodynamic approximation, scales as $\omega \sim k^2$. That implies that the term $\omega H_1(0)$ is subleading compared to $k^2 H_2(0)/\omega$ and can be ignored in the hydrodynamic approximation. Moreover, taking the near horizon limit $z \approx z_h$ implies $H_2(z) \approx 0$. With the above assumptions, the gauge field can finally be written as
\be
A_t(z)= e^{-i \omega r(z)} \frac{ \omega}{\omega+i D_c k^2 }\,.
\ee
where the diffusion constant is given by $D_c=H_2(0)$. Using that $\sqrt{-G}=G_{xx}^{\frac{d-1}{2}}\sqrt{G_{tt}G_{zz}}$, we can write the diffusion constant as
\be \label{eq-diffusionDc}
D_c=H_2(0)=\frac{\sqrt{-G}}{G_{xx}\sqrt{G_{tt}G_{zz}}}\Big|_{z_h} \int_{0}^{z_h} dz \frac{G_{tt}G_{zz}}{\sqrt{-G}}\,,
\ee
which is consistent with the results obtained in \cite{Iqbal:2008by, Starinets:2008fb}. The formula (\ref{eq-diffusionDc}) was obtained considering fluctuations of a gauge field in the probe approximation. This formula is not valid in the presence of a background gauge field. From the point of view of the boundary theory, that implies that our results are only valid for systems with zero chemical potential.

\section{Vector field bulk-to-boundary propagators in the hydrodynamic limit} \label{app-A}

In this section, we derive bulk-to-bulk propagators for a gauge field propagating in the background (\ref{eq-background}). We start by writing Maxwell-Einstein equations with a source term
\be \label{eq_gauge eom}
\partial_\mu (\sqrt{-g} g^{\mu \rho} g^{\nu \sigma} F_{\rho \sigma}) =J_\text{bulk}^{\nu}\,,
\ee
where the source satisfies the equation
\be \label{eq-continuity}
\partial_{\mu} J_\text{bulk}^{\mu} =0\,.
\ee
The gauge field can then be written as
\be
A_{\mu}({\bf r}) = \int d^{d+1}{\bf r}\, \mathcal{G}_{\mu \nu}({\bf r-r'}) J_\text{bulk}^{\nu}({\bf r'})\,,
\ee
where ${\bf r}=(r,t, x)$ and $G_{\mu \nu}({\bf r-r'})$ solves (\ref{eq_gauge eom}) with a delta function as the source. 

Following~\cite{Cheng:2021mop}, we use the gauge $A_i=0$. By writing the propagators and the sources in momentum space
\bea
\mathcal{G}_{\mu \nu}(t, x) &=&\int \frac{d\omega d^{d-1} k}{(2\pi)^{d}} \, G_{\mu \nu} (r,\omega,k) e^{-i \omega t} e^{i k \cdot x}\,, \nonumber \\
J_\text{bulk}^{\nu}(t, x) &=&\int \frac{d\omega d^{d-1}k}{(2\pi)^{d}} \, J_\text{bulk}^{\nu} (r,\omega,k) e^{-i \omega t} e^{i  k\cdot x}\, ,
\eea
we can find a solution for $\mathcal{G}_{\mu \nu} (r,\omega, k)$ in the hydrodynamic limit.
To find $\mathcal{G}_{tt}$, we require $J_\text{bulk}^{r}=0$, and $J_\text{bulk}^{t}(r,\omega,k)=\delta(r-r')$. Then (\ref{eq-continuity}) implies $J_\text{bulk}^{x}(r,\omega,k)=\frac{\omega}{k} \delta(r-r')$. The $t$ components of (\ref{eq_gauge eom}) reads
\be \label{eq-eom-t2}
\partial_r (\sqrt{-g} g^{rr} g^{tt} (\partial_r \mathcal{G}_{tt} + i \omega \mathcal{G}_{rr}))-k^2 (\sqrt{-g} g^{xx} g^{tt}\mathcal{G}_{tt})=\delta(r-r')\,.
\ee
while the $r$ component of (\ref{eq_gauge eom}) reads
\be \label{eq-eom-r2}
-i \omega (\sqrt{-g}g^{rr}g^{tt}(-i \omega \mathcal{G}_{rr} - \partial_r \mathcal{G}_{tt}) )- k^2 (\sqrt{-g}g^{xx}g^{tt} \mathcal{G}_{rr})=0\,.
\ee
By solving (\ref{eq-eom-r2}) for $\mathcal{G}_{rr}$ and substituting the result in (\ref{eq-eom-t2}), we find

\be
\partial_r \left( \frac{ \sqrt{-g}g^{rr}g^{tt}g^{xx} \partial_r \mathcal{G}_{tt}}{\omega^2 g^{tt}+k^2 g^{xx}}\right) - (\sqrt{-g}g^{xx}g^{tt} \mathcal{G}_{tt})=\frac{\delta(r-r')}{k^2}\,,
\ee
which, after some simplifications, can be written as
\be
\partial_r^2 \mathcal{G}_{tt}+ \partial_r \mathcal{G}_{tt}\, \partial_r \log \left[  \frac{ \sqrt{-g}g^{rr}g^{tt} g^{xx}}{\omega^2 g^{tt}+k^2 g^{xx}} \right]  - g_{rr}(\omega^2 g^{tt}+k^2 g^{xx}) \mathcal{G}_{tt} =\frac{g_{rr}(\omega^2 g^{tt}+k^2 g^{xx})}{\sqrt{-g}g^{tt} g^{xx}}\frac{\delta(r-r')}{k^2}\,.
\ee
A solution to the above equation satisfying in-falling boundary conditions at the horizon and Neumann boundary conditions on the boundary corresponds to the retarded bulk-to-bulk propagator in momentum space:
\be
\mathcal{G}_{tt}^{\text{ret}}(r,r',\omega, k) = -i \int dt\, d^{d-1} x e^{i \omega t -i  k \cdot x} \theta(t)\langle [A_t(r,t,x),A_t(r',0,0)]\rangle \,.
\ee
For small $\omega$ and small $k$, one can show that the solution in the near-horizon region reads~\cite{Cheng:2021mop}
\be
\mathcal{G}_{tt}^{\text{ret}}(r,r',\omega,k) \sim  \frac{\omega^2}{k^2}\frac{e^{-i \omega (r-r')}}{i \omega - D_c k^2}\,.
\ee
with $D_c$ given by (\ref{eq-diffusionDc}).

The Wightman function $i\, \mathcal{G}_{tt}^{\text{w}}=\langle A_t(r,\omega,k) A_t(r',\omega, k) \rangle$  can be written as \cite{Son:2002sd}
\be
\mathcal{G}_{tt}^\text{w}(r,r',\omega,k) = 2i \frac{e^{\omega/2T}}{e^{\omega/T}-1} \text{Im}\, \mathcal{G}_{tt}^\text{ret}(r,r',\omega,k) \,.
\ee
In position space, the above bulk-to-bulk propagators can be written as 
\bea
\mathcal{G}^\text{ret}_{tt}(r,t, x;r',t', x')&=&-i  \int \frac{d^{d-1} k}{(2\pi)^{d-1}} \frac{d \omega }{2\pi} \mathcal{G}_{tt}^\text{ret}(r,r',\omega,k) e^{-i \omega (t-t')} e^{i  k\cdot  (x-x')}  \nonumber \\
&=& -i  \int \frac{d^{d-1} k}{(2\pi)^{d-1}} \frac{d \omega }{2\pi} \frac{\omega^2}{k^2}\frac{e^{-i \omega (r+t-r'-t')}}{i \omega - D_c k^2} e^{i  k\cdot  (x-x')} \nonumber \\
&=& -i  \int \frac{d^{d-1} k}{(2\pi)^{d-1}} \frac{d \omega }{2\pi} \frac{\omega^2}{k^2}\frac{(V/V')^{- i \omega/2 \pi T}}{i \omega - D_c k^2} e^{i  k\cdot  (x-x')}  \nonumber \\
&=& i  D_c^2 \int \frac{d^{d-1} k}{(2\pi)^{d-1}} k^2 \left( \frac{V}{V'} \right)^{-D_c k^2 / 2\pi T} e^{i  k\cdot  (x-x')} \theta(V-V')\,. 
\eea
The Wightman function can be obtained as
\be
\mathcal{G}_{tt}^\text{w}(U,V, x;U',V',x') =   D_c^2 \int \frac{d^{d-1} k}{(2\pi)^{d-1}} \frac{k^2 \, e^{i  k\cdot  (x-x')}}{2 \sin \left( \frac{D_c k^2}{2 T} \right)} \left[ \left(\frac{V}{V'} \right)^{-\frac{D_c k^2}{  2\pi T}} -\left( \frac{U}{U'} \right)^{\frac{D_c k^2}{  2\pi T}} \right] \,.
\ee
 The corresponding bulk-to-boundary propagators can be obtained from the above expressions by taking one of the bulk points to the boundary. Denoting the bulk point as $(U=0,V, x)$ and the boundary point as $(t_1,x_1)$, we can write the retarded bulk-to-boundary propagator as
 \be
 K^\text{ret}_{tt}(V, x;t_1, x_1) = i\,  D_c^2 \int \frac{d^{d-1} k}{(2\pi)^{d-1}} k^2 \frac{e^{i  k \cdot  (x-x_1)} }{\left( V \,e^{-2 \pi T t_1} \right)^{\frac{D_c k^2}{  2\pi T}}} \theta \left(V-e^{2 \pi T t_1} \right)\,,
 \ee
 and the Wightman bulk-to-boundary propagator as
 \be
 K^\text{w}_{tt}(V, x;t_1,x_1)  =  D_c^2 \int \frac{d^{d-1} k}{(2\pi)^{d-1}} \frac{k^2}{2 \sin \left( \frac{D_c k^2}{2 T} \right)} \frac{e^{i  k\cdot (x-x_1)} }{\left( V\, e^{-2\pi T t_1} \right)^{\frac{D_c k^2}{  2\pi T}}} \,.
 \ee
 where we used that $r(z \rightarrow 0)=0$ and consequently $V \rightarrow e^{2 \pi T t_1}$ as $z \rightarrow 0$.

In the calculation of $\Delta U$, we need $ K^\text{w}_{Vt}$ and $ K^\text{ret}_{Vt}$, which can easily be obtained from $K^\text{w}_{tt}$ and $ K^\text{ret}_{tt}$. Since $A_V=\frac{A_t+A_r}{2\pi T V}$, we can write $\langle A_V J_t \rangle = \frac{1}{2\pi T V} \langle (A_t+A_r) J_t \rangle$. This can be simplified even more in the hydrodynamic limit, in which $A_r = A_t$. Using this last equality, we find $\langle A_V J_t \rangle = \frac{1}{\pi T V} \langle A_t J_t \rangle$. With the above considerations, we find
\bea
 K^\text{ret}_{Vt}(V, x;t_1, x_1)&=& 2 i  D_c\,T\, \theta(V-e^{2\pi T t_1}) \, \partial_V \int \frac{d^{d-1} k}{(2\pi)^{d-1}} \frac{e^{i k \cdot (x-x_1)}}{\left( V/V_1 \right)^{\frac{D_c k^2}{2\pi T}}}  \,, \label{eq-K1}\\
 K^\text{w}_{Vt}(V,x;t_1, x_1)  &=&  T\,D_c\,  \partial_V \int \frac{d^{d-1} k}{(2\pi)^{d-1}} \frac{1}{2 \sin \left( \frac{D_c k^2}{2 T} \right)} \frac{e^{i  k\cdot  (x-x_1)} }{\left( V/V_1 \right)^{\frac{D_c k^2}{  2\pi T}}} \label{eq-K2}\,.
\eea
where $V_1=e^{-2\pi T t_1}$, and an overall factor of $T$ was included for dimensional reasons.
These formulas were derived in the hydrodynamic limit, which means they are only well-defined near the black hole horizon. Since we consider the expectation value of stress energy tensor in the near horizon region, the large time separation modes ($V/V_1 \gg 1$) give a dominant contribution to it.

Also, the stress tensor $\langle T_{VV} \rangle $ computed from the above propagators is independent on the insertion time $t_0$. This happens because $\langle T_{VV} \rangle $ is proportional to\footnote{Here we omit the other coordinates for simplicity.} $K^\text{ret}(V,t_0) K^\text{w}(V,-t_0)$ and the $t_0$ dependence cancels out. Consequently, the expression for the averaged null energy $\int_{V_0}^{\infty} dV \langle T_{VV} \rangle$ derived using (\ref{eq-K1}) and (\ref{eq-K2}) is only valid at late times. To fix this problem and obtain a result for ANE that depends on the insertion time, we used the propagators proposed by Swingle and Cheng in \cite{Cheng:2021mop}:

\bea
 K^\text{ret}_{Vt}(V, x;t_1, x_1)&=& 2 i T\, D_c\,\, \theta(V-e^{2\pi T t_1}) \, \partial_V \int \frac{d^{d-1} k}{(2\pi)^{d-1}} \frac{e^{i k \cdot (x-x_1)}}{\left( V e^{-2\pi T t_1} -1\right)^{\frac{D_c k^2}{2\pi T}}}  \,,\\
 K^\text{w}_{Vt}(V,x;t_1, x_1)  &=& T\, D_c\,  \partial_V \int \frac{d^{d-1} k}{(2\pi)^{d-1}} \frac{1}{2 \sin \left( \frac{D_c k^2}{2 T} \right)} \frac{e^{i  k\cdot  (x-x_1)} }{\left(1- V\, e^{-2\pi T t_1} \right)^{\frac{D_c k^2}{  2\pi T}}} \,.
\eea
According to \cite{Cheng:2021mop}, the above propagators have qualitatively the same behavior as (\ref{eq-K1}) and (\ref{eq-K2})  in the late time limit and have the desired analytic properties connecting $K^\text{ret}$ and $K^\text{w}$. They allow us to study the dependence of ANE for any value of the insertion time, which is more physical, since there is nothing special about $t_0=0$.

\bibliographystyle{JHEP}

\end{document}